\documentclass[11pt]{article}
\usepackage{aas_macros,amsmath,amssymb,comment,cite,esint,graphicx,mathtools}
\usepackage[margin=.8in,letterpaper]{geometry}
\usepackage[colorlinks=true]{hyperref}
\usepackage[affil-it]{authblk}
\usepackage{subcaption}
\usepackage[utf8]{inputenc}
\usepackage{mathrsfs}
\usepackage{appendix}
\usepackage{amssymb}
\usepackage{float}
\usepackage{color}
\usepackage{cite}
\usepackage{hyperref}
\usepackage{longtable}
\hypersetup{pageanchor=false}
\usepackage{indentfirst}
\usepackage{url}
\usepackage{float}
\usepackage{caption}
\usepackage[numbers,square,comma,sort&compress,merge]{natbib}
\usepackage{esint}
\usepackage{overpic}
\usepackage{graphicx}
\usepackage{epsf,amsmath,bbold,amsfonts,stmaryrd}
\usepackage{textcomp}
\usepackage{ulem}
\usepackage{tikz}
\usepackage{multirow}

\numberwithin{equation}{section}
\let\originalleft\left
\let\originalright\right
\renewcommand{\left}{\mathopen{}\mathclose\bgroup\originalleft}
\renewcommand{\right}{\aftergroup\egroup\originalright}
\mathcode`\*="8000
{\catcode`\*=\active\gdef*{\mathclose{}\,\mathopen{}}}

\title{Generally relativistic description of fast magnetic reconnection induced by thermal electromotive force}

\author{
Ye Shen$^{1}$\thanks{E-mail: shenye199594@stu.pku.edu.cn}~
\\
$^{1}$School of Physics, Peking University, No.5 Yiheyuan Rd, Beijing
		100871, P.R. China
}

\begin{document}

\maketitle

\begin{abstract}
    \vspace{5mm}
Many theoretical models were come up with to figure out the properties of magnetic reconnection process, among which the Sweet-Parker model is the most famous since it describes the magnetic reconnection in a concise way. However, the low reconnection rate expected by this model is generally not available in most astrophysical systems, which motivates people to seek fast reconnection models. Under the scheme of generalized magnetohydrodynamics (MHD) for pair plasma, a fast magnetic reconnection model was established, in which the thermal electromotive force plays a key role to remarkably increase the reconnection rate. In this work, I would like to extend the discussions in my previous work, about the generally relativistic description of Sweet-Parker model, to the description of fast magnetic reconnection induced by thermal electromotive force. I will revisit the fast reconnection model briefly to initialize my discussions and show how the thermal electromotive force impacts the reconnection rate. Next, some basic setups will be exhibited before discussing specific examples about how the properties of fast magnetic reconnection are modified by gravitational effect or in observations. Results in this work consolidate my opinion reiterated in my previous work that properties of magnetic reconnection would never be modified by gravitational effect significantly if the magnetic reconnection process occurs in a local scale while the modifications of properties could not be neglected when the process is detected by an observer who is moving with respect to the laboratory, in the rest frame of which the magnetic reconnection occurs.
\end{abstract}

\begin{keywords}
Fast Magnetic Reconnection -- General Relativity -- Generalized MHD
\end{keywords}

\newpage
\baselineskip 18pt

\tableofcontents  

\section{Introduction}
\label{sec:intro}

In order to explain the solar flares \cite{substorm,2021SSRv..217...66Z}, people came up with the magnetic reconnection process, in which two branches of anti-parallel magnetic field lines touch each other and the configuration of magnetic field is reconstructed in a local diffusion region. From then on, as a physical process that is capable of releasing magnetic energy efficiently in astrophysical scenario, magnetic reconnection draws great attention such that a big amount of attempts on its analytical models and numerical simulations were made.

As the first analytical model, the Sweet-Parker model described the magnetic process under the scheme of magnetohydrodynamics (MHD) \cite{SP1,SP2,Lyubarsky2006}. Some properties, such as the outflow speed equaling the local Alfv{\'e}n velocity, have already been supported by numerical simulations \cite{Comisso:2023ygd,Sironi:2014jfa}. However, it is fatal that the reconnection rate expected by Sweet-Parker model is extremely low, which is in contrast to the observations of most astrophysical systems \cite{yamada2009,1973ApJ...180..247P} and motivates people to explore fast reconnection models. Petschek proposed the concept of local reconnection rate for the first time. Different from the Sweet-Parker model, Petschek argued that the inflow plasma upstream is not required to be all ejected from two sides of the narrow current sheet so that a much higher reconnection rate could be obtained as a consequence \cite{Petschek}. However, the analyses of the so called Petchek model are not concise and its self-consistency is still in doubt \cite{Biskamp1986,Sato1979}. 

Collisionless reconnection was initially developed in numerical works by inducing Hall effect, which was once regarded to be critical for generating fast reconnection rate \cite{2008PhPl...15d2306D,2006PhPl...13e2119Y}. However, further numerical studies showed that fast reconnection rate could still be produced when the Hall term vanishes or is inactive \cite{Bessho:2005zz,Swisdak:2008xn,2015PhPl...22a0701S}. A simple analytical model was proposed under the MHD scheme in Ref.~\cite{Liu2017}, which import a geometric index to control the shape of diffusion region and the configuration of magnetic field in local scale. In this model, high reconnection rates would be got by choosing suitable values of the geometric index and it is independent of the mechanism of dissipation. However, too many assumptions make it to be just a toy model that is concise but has a long distance away from the reality. In Ref.~\cite{CA2014}, the Sweet-Parker and Petschek configurations were extended to the generalized MHD scheme \cite{Koide2009,Koide2010}, in which two kinds of magnetofluid carrying positive and negative charges dissipate separately. As another fast reconnection model, it indicated that the thermal electromotive force, which does not exist in the standard MHD scheme, plays a key role. A high reconnection rate could be obtained even though the anti-parallel magnetic field lines approach each other in the way as the Sweet-Parker model presumed.

Since it has been widely accepted that the magnetic reconnection occurs in the accretion flow around compact objects \cite{Jia:2023iup,Davelaar:2023dhl,Aimar2023,Ripperda2020,Yuan2024-1,Yuan2024-2}, works related to the magnetic reconnection process near a black hole become more and more attractive in recent years, such as the energy extraction via magnetic reconnection from a rotating black hole \cite{KA2008,CA2021,Work0,Work1}, which motivate people to explore the description of magnetic reconnection process in curved spacetime. A generally relativistic (GR) description of Sweet-Parker model was made in Ref.~\cite{CA2017}, where the magnetic reconnection was considered to occur in the rest frame of the zero-angular-momentum observers (ZAMOs). The discussions in Ref.~\cite{CA2017} have been broadened to the rest frame of fluid moving on the equatorial plane in Ref.~\cite{Fan:2024fcy}, where the Petschek model was briefly discussed as well. The GR description of a fast reconnection model in pair plasma, adopting Sweet-Parker configuration under the generalized MHD scheme, was introduced in Ref.~\cite{CA2018,Fan:2024rsa}. However, as exploratory works, there are some problems pending to be solved. For example, in their works, equations in standard or generalized MHD scheme were all projected onto the rest frame of ZAMOs, which is cumbersome and unnecessary, especially when discussing the process occurring in the fluid's rest frame. Moreover, it is inevitable to determine the rest frame of laboratory, where the process occurs, and the rest frame of observer, in which the process is detected, before describing a physical process in GR. Though the laboratories were determined in Ref.~\cite{CA2017,CA2018,Fan:2024fcy}, the observer was not under their consideration. Additionally, in Sweet-Parker configuration, derivatives in the equations should be substituted by finite differences in order to result in concise relations between physical quantities \cite{Lyubarsky2006}. However, these substitutions seems to be chosen in a wrong way in Ref.~\cite{CA2017,CA2018,Fan:2024fcy}. Based on this incorrect choice, they provided the modifications on properties of magnetic reconnection induced by spacetime curvature.

In Ref.~\cite{GRMR1}, I tried to figure out the GR description of the Sweet-Parker model in a different way. I reorganized the calculations of Sweet-Parker model in special relativity (SR) into seven steps, whose corresponding GR forms were proposed after determing the laboratory, observer and how the current sheet is posited. The modifications on the properties of magnetic reconnection coming from gravitational effect were discussed by choosing the substitutions different from the choices in Ref.~\cite{CA2017,CA2018,Fan:2024fcy}. As a main conclusion, it was reiterated in Ref.~\cite{GRMR1} that the modifications induced by gravitational effect should be infinitesimal if the local scale, within which the magnetic reconnection occurs, is tiny compared to the horizon scale of the central black hole. Furthermore, I discussed the situation when the outflow speed and reconnection rate are detected by an observer who is moving with respect to the laboratory. The results indicated that properties of magnetic reconnection process would be modified significantly when it is not detected in the rest frame of laboratory.

As a sequel, in this work, I would like to extend my discussions in Ref.~\cite{GRMR1} to the fast reconnection model in the pair plasma with the Sweet-Parker configuration adopted, for which the equations of generalized MHD should be utilized. I will recap the fast reconnection induced by thermal electromotive force in pair plasma in SR briefly and show how the thermal electromotive force increases the reconnection rate. The basic setups in Ref.~\cite{GRMR1} will be revisited and applied to the generalized MHD scheme for describing the fast magnetic reconnection in GR. Then I will discuss the modifications on the properties of reconnection caused by gravitational effect and observation. Similar to the conclusions in Ref.~\cite{GRMR1}, results in this work indicate that the modifications caused by gravitational field should be infinitesimal in local scale, which opposes the conclusions in Ref.~\cite{CA2017,CA2018,Fan:2024fcy}, but the properties would be significantly modified if it is not detected in the rest frame of laboratory.

The outline of the paper is as follows. In Sect.~\ref{sec:SR}, equations of generalized MHD and the fast reconnection model induced by thermal electromotive force is revisited. The GR scenario is discussed in Sect.~\ref{sec:general}, including the equations of generalized MHD, configuration of electromagnetic field in the local scale and the effects of observations. The modifications on the property of reconnection rate generated by gravitational effect and observation are discussed in Sect.~\ref{sec:example} by discussing some specific examples. A brief summary is made in Sect.~\ref{sec:sum}. How the equations in generalized MHD are derived is shown in Appen.~\ref{sec:gMHD}. In Appen.~\ref{sec:resistivity}, the relation between the electrical resistivity and the collision rate in pair plasma are shown. Some discussions and results in Ref.~\cite{GRMR1}, which would be useful in this work, are listed in Appen.~\ref{sec:SP}. The units for $G=M=c=1$ are adopted throughout the paper, where $M$ is the mass of central black hole.

\section{SR description}
\label{sec:SR}

\begin{figure}
    \centering
    \includegraphics[width=\textwidth]{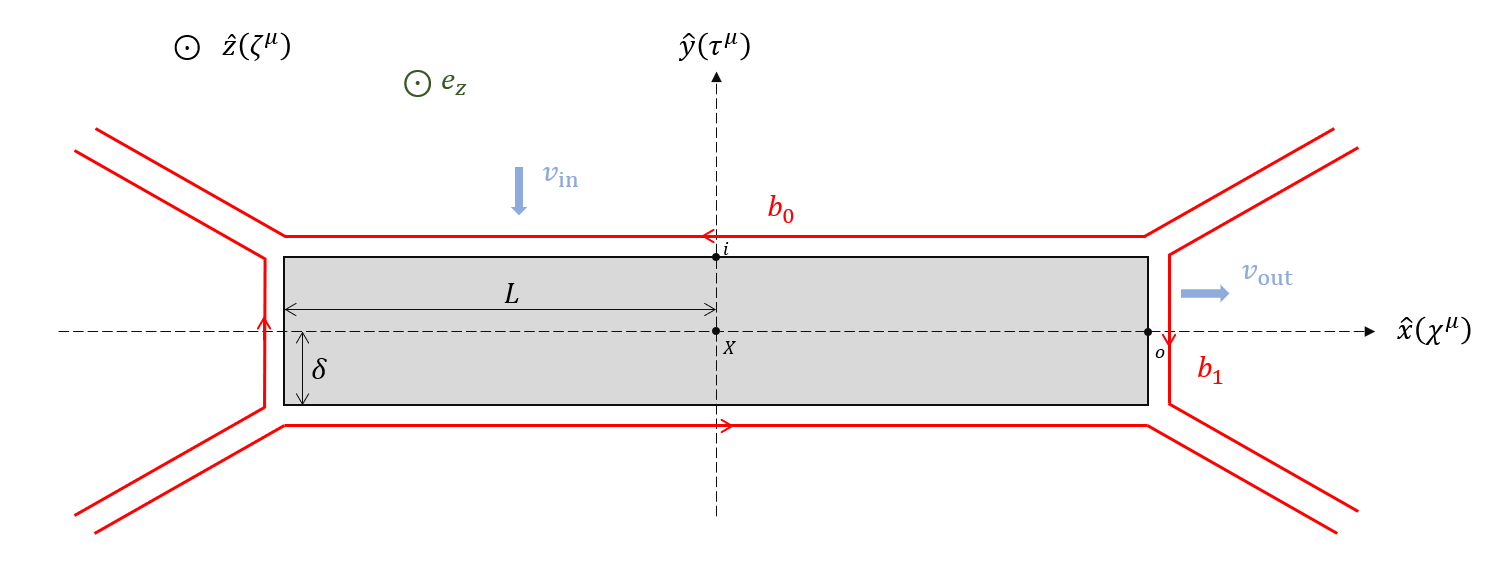}
    \caption{Fig.~1 in Ref.~\cite{GRMR1}, schematic diagram of Sweet-Paker configuration, with red lines representing the magnetic field and the grey square representing the current sheet. Here $X$ denotes the center of current sheet while $i$ and $o$ denotes the surfaces of current sheet upstream and downstream.}
    \label{fig:pic}
\end{figure}

Equations of the generalized MHD scheme, consisting of two kinds of magnetofluid carrying positive and negative charges respectively, were derived in Ref.~\cite{Koide2009}.  The mass conservation and Maxwell's equations in generalized MHD scheme are identical to those in the standard MHD scheme:
\begin{equation}
    \begin{split}
        \text{\bf mass conservation:}&~\nabla\cdot\left(\gamma\Vec{v}\right)=0  \\
        \text{\bf Maxwell's equations:}&~\nabla\cdot\Vec{E}=0~,~\nabla\times\Vec{E}=0~,~\nabla\cdot\Vec{B}=0~,~\nabla\times\Vec{B}=\Vec{J} \\
    \end{split}
    \label{eq:MHD1}
\end{equation}
where $\Vec{v}$ is averaged velocity weighted by mass, obeying $\gamma\Vec{v}=\left(m_+n_+\gamma_+\Vec{v}_++m_-n_-\gamma_-\Vec{v}_-\right)/\rho$, with $m_{\pm}$ the masses of two kinds of particles and $\rho=m_+n_++m_-n_-$ the total mass density. The subscript "$\pm$" denotes the positively and negatively charged particles in plasma respectively. The current density could be expressed as $\Vec{J}=e\left(n_+\gamma_+\Vec{v}_+-n_-\gamma_-\Vec{v}_-\right)$, with $e$ the amount of charge in one particle. Here, we adopt the quasi-stationary condition
\begin{equation}              
    \partial_t q \sim 0
    \label{eq:steady}
\end{equation}
with $q$ an arbitrary physical quantity, and hopes the plasma is neutral approximately such that: $\gamma_+n_+ \approx \gamma_-n_-$. The particle number densities, pressures and enthalpies of the magnetofluid are assumed to reach uniformity in the local scale, within which the magnetic reconnection occurs.

Since the positively and negatively charged particles flows separately, both the momentum conservation and Ohm's law should be generalized, for which at least four additional effects should be considered. They are Hall effect, current inertia, thermal electromotive force and the thermal exchange between positively and negatively charged fluid \cite{BirnBook}. The Hall effect takes place when one kind of particles could drift with magnetic field while another one could not. The current inertia and thermal electromitive force reflect how the current in the plasma could stand up to the perturbation coming from electromagnetic field or other kinds of effects. In the following, for simplicity, we neglect the thermal exchange and consider the case of pair plasma and ion-electron plasma. One can read Appen.~\ref{sec:gMHD} to figure out how the momentum conservation and Ohm's law are derived from two-fluid's equations of motion in generalized MHD in detail.

In the pair plasma consisting of positrons and electrons such that $m_+\approx m_-$, momentum conservation and Ohm's law in generalized MHD scheme read:
\begin{equation}
    \begin{split}
        \text{\bf momentum conservation:}&~\nabla\cdot\left(\omega\gamma^2\Vec{v}\Vec{v}+\frac{\omega}{4{\frak q}^2}\Vec{J}\Vec{J}\right)+\nabla p=\Vec{J}\times\Vec{B}  \\
        \text{\bf Ohm's law:}&~\Vec{E}+\Vec{v}\times\Vec{B}=\eta\Vec{J}+\frac{\omega}{4{\frak q}^2}\nabla\cdot\left[\gamma\left(\Vec{v}\Vec{J}+\Vec{J}\Vec{v}\right)\right] \\
    \end{split}
    \label{eq:MHD2-limit1}
\end{equation}
where $\omega=n^2\left(\omega_+/n_+^2+\omega_-/n_-^2\right)$ is the averaged proper enthalpy, $p=p_++p_-$ is the total pressure and ${\frak q}=ne$ is called the latent charge density, with $n=(n_+m_++n_-m_-)/(m_++m_-)$ the averaged particle number density. We adopt $p_+\approx p_-$ for the sake of simplicity in Eq.~\eqref{eq:MHD2-limit1}, align with the simplification utilized in Ref.~\cite{CA2014}. Another term $\eta Q\gamma\Vec{v}(1+\Theta)$ should be added on the right side of Ohm's law if thermal exchange should be considered, where $\Theta$ corresponds to the friction between two kinds of fluid while $Q=e\left(\gamma_+n_+-\gamma_-n_-\right)$ represents the charge density of plasma. The term $\frac{\gamma\omega}{4{\frak q}^2}\left(\Vec{v}\Vec{J}+\Vec{J}\Vec{v}\right)$ corresponds to the thermal electromotive force (called thermal-inertial term in Ref.~\cite{CA2014}) and $\frac{\omega}{4{\frak q}^2}\Vec{J}\Vec{J}$ is the current inertia \cite{Koide2009}. In pair plasma, since the the inertia of positrons and electrons are almost equal, it is barely realized that one kind of particles could drift with the magnetic field while another one could not. Subsequently, the Hall effect could be neglected.

In the ion-electron plasma where $m_+\gg m_-$, momentum conservation and Ohm's law in generalized MHD scheme read: 
\begin{equation}
    \begin{split}
        \text{\bf momentum conservation:}&~\nabla\cdot\left(\omega\gamma^2\Vec{v}\Vec{v}\right)+\nabla p=\Vec{J}\times\Vec{B}  \\
        \text{\bf Ohm's law:}&~\Vec{E}+\Vec{v}\times\Vec{B}=\eta\Vec{J}+\frac{1}{{\frak q}}\left(\Vec{J}\times\Vec{B}-\frac{1}{2}\nabla p_-\right) \\
    \end{split}
    \label{eq:MHD2-limit2}
\end{equation}
The term $\Vec{J}\times\Vec{B}/{\frak q}$ depicts the Hall effects, which was introduced at first when people considered the collisionless magnetic reconnection in numerical simulations \cite{yamada2009,2006PhPl...13e2119Y,2008PhPl...15d2306D}. In the ion-electron plasma, since the inertia of electrons is much less than that of ions, the Hall effect becomes important. While on the contrary, since the motion of electrons could be vastly changed by electromagnetic field or other effects in the case that the motion of ions is barely affected, the current in the ion-electron plasma is less resistive under perturbations. Hence, the current inertia and thermal electromotive force could be neglected.

Now let us consider the magnetic reconnection process in pair plasma or ion-electron plasma. We adopt the Sweet-Parker configuration hereafter for simplicity, whose schematic diagram is plotted in Fig.~\ref{fig:pic}. The current sheet (called magnetic diffusion region as well), within which the current is allowed to exist only, is posited in $\hat{x}-\hat{y}$ plane. The current sheet is assumed to be a thin rectangle whose length $L$ is much longer than the width $\delta$. The reconnection point, surfaces of current sheet upstream and downstream are denoted by $X$, $i$ and $o$ respectively. Magnetic strengths upstream and downstream are set to be $\Vec{B}\Big|_i=b_0\hat{x}$ and $\Vec{B}\Big|_o=b_1\hat{y}$ while the electric field directs along $\hat{z}$ axis. Plasma flows into the current sheet along $\hat{y}$ upstream with speed $v_{\rm in}$ while flows out along $\hat{x}$ downstream with speed $v_{\rm out}$. In order to get concise relations between physical quantities, derivatives in the equations could be substituded by finite differences in the form of \cite{Lyubarsky2006}
\begin{equation}
    \partial_xq\sim \frac{q\big|_o-q\big|_X}{L}~~~,~~~\partial_yq\sim\frac{q\big|_X-q\big|_i}{\delta}
    \label{eq:L-delta-SR}
\end{equation}
In Ref.~\cite{GRMR1}, calculations of Sweet-Parker model were reorganized into seven steps, which are listed in Appen.~\ref{sec:SP}. Since the generalized MHD scheme only changes the momentum conservation and Ohm's law compared to the standard one, some relations and properties in Sweet-Parker model are still useful. We only need to reconsider the steps corresponding to the momentum conservation and Ohm's law, which are the third, sixth and seventh steps (see Appen.~\ref{sec:SP}). 

For the pair plasma, Ohm's law upstream and inside the current sheet gives:
\begin{equation}
    \begin{split}
        \text{upstream:}&~e_z+v_{\rm in}b_0=0 \\
        \text{current sheet:}&~e_z\simeq (\eta+\Lambda)j_z
    \end{split}
    \label{eq:s3-1}
\end{equation}
Here $\Lambda$ is called the thermal-inertial parameter, defined to be \cite{CA2014}:
\begin{equation}
    \Lambda=\frac{\omega}{4\frak{q}^2}\nabla\cdot\left[\gamma\left(\Vec{v}\Vec{J}+\Vec{J}\Vec{v}\right)\right] \sim \frac{\omega\gamma_{\rm out}v_{\rm out}}{4\frak{q}^2L}
    \label{eq:Lambda-1}
\end{equation}
It is induced by thermal electromotive force. The approximation $v_{\rm in}\ll v_{\rm out}$ (from Eq.~\eqref{eq:step5-res}) is applied in Eq.~\eqref{eq:Lambda-1} for the final expression. Eq.~\eqref{eq:s3-1} indicates that the thermal-inertial parameter plays a role of analogous resistivity. Since $\Vec{J}$ has $\hat{z}$ component only (one could figured it out from the symmetry of the configuration as well), the current inertia would not appear in the following calculations. Then the momentum conservation returns back to the form in standard MHD. Along $\hat{y}$ axis, the momentum conservation gives:
\begin{equation}
    \partial_y\left(\omega\gamma^2V_y^2\right)+\partial_yp=J_zB_x~~~\rightarrow~~~p_X\simeq \frac{b_0^2}{2}
\end{equation}
for $V_y\ll 1$. While along $\hat{x}$ axis it gives:
\begin{equation}
    \partial_x\left(\omega\gamma^2V_x^2\right)+\partial_xp=J_zB_y~~~\rightarrow~~~\frac{1}{L}\left(\omega\gamma_{\rm out}^2v_{\rm out}^2-p_X\right)\simeq \frac{b_0^2}{2}
\end{equation}
with Eq.~\eqref{eq:step4-res} applied in the equation on the right side. So we have:
\begin{equation}
    v_{\rm out}\simeq v_{\rm A}=\sqrt{\frac{\sigma_0}{1+\sigma_0}}
    \label{eq:vout-SR}
\end{equation}
where $\sigma_0$ represents the local magnetization of plasma while $v_{\rm A}$ represents the local Alfv{\'e}n velocity. One can see that, for pair plasma, the outflow speed deduced from generalized MHD scheme still equals the local Alfv{\'e}n velocity, obeying the property we got in Sweet-Parker model. However, the reconnection rate becomes:
\begin{equation}
    R\equiv\frac{v_{\rm in}}{v_{\rm A}}\simeq \frac{\delta}{L}\simeq \frac{\eta+\Lambda}{\delta v_{\rm A}} \simeq \sqrt{\frac{1}{S}+\frac{\omega\sqrt{1+\sigma_0}}{4{\frak q}^2L^2}}
    \label{eq:R-1}
\end{equation}
where $S$ is the Lundquist number that is very high in most astrophysical systems \cite{yamada2009}. The existence of thermal-inertial parameter increases the reconnection rate. As mentioned above, the thermal-intertial parameter acts just like the resistivity, which is also called the magnetic diffusivity in another point of view. In this sense, the magnetic diffuses faster from the current sheet when the thermal electromotive force exists, and hence the reconnection rate increases. 

When the thermal electromotive force becomes dorminant for the diffusion of magnetic field, we generally have $\Lambda\gg \eta$. In the pair plasma, the magnetic diffusivity obeys $\eta\simeq \rho \bar{\nu}_{\rm pe}/\frak{q}^2$, where $\bar{\nu}_{\rm pe}$ is the rate of collision between positrons and electrons (read Appen.~\ref{sec:resistivity} for the details). Thus, the dormination of thermal electromotive force in the magnetic reconnection process requires:
\begin{equation}
    \Lambda\gg\eta~~~\rightarrow~~~\lambda_{\rm in}\equiv\frac{v_{\rm in}}{\bar{\nu}_{\rm pe}}\gg \delta
    \label{eq:collisionless}
\end{equation}
where the relation between $v_{\rm in}$ and $v_{\rm out}$ is utilized. We adopted $\rho \simeq \omega$ and $\gamma_{\rm in}\simeq 1$ in Eq.~\eqref{eq:collisionless} for simplicity. The $\lambda_{\rm in}$ could be regarded as the mean free path of positron and electron upstream with respect to the collision between them. Eq.~\eqref{eq:collisionless} indicates that collision between positrons and electrons could be ignored within the current sheet when the thermal electromotive force plays an important role. Subsequently, for a magnetic reconnection process occurring with high reconnection rate in the pair plasma with large Lundquist number, it means the positrons and electrons are fully decoupled locally. On the other hand, when the thermal electromotive force gives a great impact in pair plasma, its divergent should be larger than or at the scale of the gradient of pressure, which requires $\big|\Vec{v}_+-\Vec{v}_-\big|\gtrsim p/\omega \sim c_s$. Here $c_s$ represents the sound speed. It means, when the thermal electromotive force becomes important, the relative drift between positrons and electrons should be supersonic.

For electron-ion plasma, only Ohm's law changes. However, for the Hall term, in the Sweet-Parker configuration, it vanishes upstream because there is no current outside the current sheet while it vanishes inside the current sheet because of the vanishing magnetic field. So the Hall term has no effect actually. From another point of view, magnetic reconnection under the domination of Hall effect could never occur in the Sweet-Parker configuration. The second term in the parathesis relates to the pressure of electrons. In the scenario of two-temperature accretion disk, temperature of electrons are generally believed to be much lower than that of ions \cite{Vincent:2022fwj,Hou:2023bep,Zhang:2024lsf}. If the fluid consisting of ions and electrons obeys the ideal gas law, we would subsequently have $p\simeq p_+\gg p_-$, which means the influence of electrons' pressure is actually negligible. In this sense, for the electron-ion plasma, the generalized MHD equations are all identical to the standard MHD equations approximately, so as their resultant properties of magnetic reconnection, under the Sweet-Parker configuration. Hence, in the following discussions, let us focus on the pair plasma only.

\section{GR description: general discussions}
\label{sec:general}

\subsection{Basic setups}
\label{sec:basic}

Basic equations of generalized MHD for the pair plasma in GR are \cite{Koide2009,Koide2010}:
\begin{equation}
    \begin{split}
        \text{\bf{ mass conservation:}}&~~\nabla_{\mu}(\rho u^{\mu})=0 \\
        \text{\bf{ energy-momentum~conservation:}}&~~\nabla_{\nu}\left(T^{\mu\nu}+\frac{\omega}{4{\frak q}^2}J^{\mu}J^{\nu}\right)=F^{\mu\nu}J_{\nu} \\
        \text{\bf{ Ohm's~law:}}&~~F^{\mu\nu}u_{\nu}=\eta\left[J^{\mu}+J_{\nu}u^{\nu}u^{\mu}(1+\Theta)\right]+\nabla_{\nu}\left[\frac{\omega}{2{\frak q}^2}u^{(\mu}J^{\nu)}\right] \\
        \text{\bf{ Maxwell's~equations:}}&~~\nabla_{\nu}F^{\mu\nu}=J^{\mu}~~,~~\nabla_{\nu}F^{\ast\mu\nu}=0
    \end{split}
    \label{eq:GR-MHD1}
\end{equation}
Here $T^{\mu\nu}=\omega u^{\mu}u^{\nu}+pg^{\mu\nu}$ is the stress energy tensor of magnetofluid (containing positively and negatively charged ones), $\frac{\omega}{2{\frak q}^2}u^{(\mu}J^{\nu)}=\frac{\omega}{4{\frak q}^2}\left(u^{\mu}J^{\nu}+u^{\nu}J^{\mu}\right)$ corresponds to the thermal electromotive force while $\frac{\omega}{4{\frak q}^2}J^{\mu}J^{\mu}$ is the current inertia. The averaged quantities are defined similarly to those in SR. For example, the averaged velocity obeys $u^{\mu}=\left(m_+n_+u_+^{\mu}+m_-n_-u_-^{\mu}\right)/\rho$ with $\rho=m_+n_++m_-n_-$. In the local scale, $\rho$, $\omega$, $\eta$ and ${\frak q}$ reaches uniformity. The local scale for a magnetic reconnection occurring near a BH just means $\delta\ll L\ll r_g$ (in Sweet-Parker configuration), where $r_g\equiv 1$ is the gravitational radius. The plasma is considered to be neutral such that $J_{\alpha}u^{\alpha}=0$ is always satisfied.

The covariant divergences of an arbitrary vector $A^{\mu}$ and of an arbitrary two dimensional tensor $G^{\mu\nu}$ satisfy:
\begin{equation}
    \begin{split}
        \nabla_{\mu}A^{\mu}=&\frac{1}{\sqrt{-g}}\partial_{\mu}\left(\sqrt{-g}A^{\mu}\right) \\
        \nabla_{\mu}G^{\mu\nu}=&\frac{1}{\sqrt{-g}}\partial_{\mu}\left(\sqrt{-g}G^{\mu\nu}\right)+\Gamma^{\mu}_{\kappa\lambda}G^{\lambda\kappa}
    \end{split}
    \label{eq:divergence}
\end{equation}
with $\Gamma^{\mu}_{\kappa\lambda}$ the Christoffol symbols (affine connection). Applying Eq.~\eqref{eq:divergence} and all the conditions discussed above, equations of generalized MHD for the pair plasma in GR could be written as:
\begin{equation}
    \begin{split}
        &\text{\bf{ mass conservation:}} \\
        &~~~~\partial_{\mu}(\sqrt{-g} u^{\mu})=0 \\
        &\text{\bf{ energy-momentum~conservation:}} \\
        &~~~~\frac{1}{\sqrt{-g}}\partial_{\nu}\left(\sqrt{-g}\omega u^{\mu}u^{\nu}+\frac{\sqrt{-g}\omega}{4{\frak q}^2}J^{\mu}J^{\nu}\right)=-\partial^{\mu}p+F^{\mu\nu}J_{\nu}-\omega\Gamma^{\mu}_{\kappa\lambda}\left(u^{\kappa}u^{\lambda}+\frac{1}{4{\frak q}^2}J^{\kappa}J^{\lambda}\right) \\
        &\text{\bf{ Ohm's~law:}} \\
        &~~~~F^{\mu\nu}u_{\nu}=\eta J^{\mu}+\frac{1}{\sqrt{-g}}\partial_{\nu}\left[\frac{\sqrt{-g}\omega}{2{\frak q}^2}u^{(\mu}J^{\nu)}\right]+\frac{\omega}{2{\frak q}^2}\Gamma^{\mu}_{\kappa\lambda}u^{(\kappa}J^{\lambda)} \\
        &\text{\bf{ Maxwell's~equations:}} \\
        &~~~~\partial_{\nu}\left(\sqrt{-g}F^{\mu\nu}\right)=\sqrt{-g}J^{\mu}~~
                                   ,~~\partial_{\nu}\left(\sqrt{-g}F^{\ast\mu\nu}\right)=0
    \end{split}
    \label{eq:GR-MHD2}
\end{equation}
The anti-symmetries of Maxwell's tensors were used in Eq.~\eqref{eq:GR-MHD2}. Comparing to the equations of standard MHD scheme, only Ohm's law and energy-momentum conservation are different, where the thermal electromotive force and current inertia are taken into account. 

In order to describe the magnetic reconnection process in GR, it is indispensable to figure out the rest frame of laboratory, in which the magnetic reconnection occurs, and the rest frame of observer, in which all the properties of magnetic reconnection are detected. Furthermore, if the Sweet-Parker approach is adopted, one should make it clear how the current sheet is posited. In this sense, we define \cite{GRMR1}:
\begin{equation}
    \begin{split}
        k^{\mu}&  \text{: 4-velocity of laboratory}\\
        n^{\mu}&  \text{: 4-velocity of observer}\\
        \chi^{\mu}&  \text{: spacelike unit vector along which the magnetic field is posited upstream}\\
        \tau^{\mu}&  \text{: spacelike unit vector along which the magnetic field is posited downstream}\\
        \zeta^{\mu}&  \text{: spacelike unit vector along which the electric field is posited}
    \end{split}
    \label{eq:tetrad}
\end{equation}
Since the magnetic reconnection occurs in the rest frame of laboratory, the current sheet should be posited on a spacelike supersurface perpendicular to $k^{\mu}$. In other words, it is required that $k^{\mu}\chi_{\mu}=k^{\mu}\tau_{\mu}=k^{\mu}\zeta_{\mu}=0$. Moreover, in the Sweet-Parker configuration, the current sheet is set to be a rectangle, which indicates that the three spacelike unit vectors defined above are perpendicular to each other. Actually, $\chi^{\mu}$, $\tau^{\mu}$ and $\zeta^{\mu}$ maps to the $\hat{x}$, $\hat{y}$ and $\hat{z}$ axes in SR, as shown in Fig.~\ref{fig:pic}.

After determining the rest frame of laboratory, the Maxwell's tensors could be written as
\begin{equation}
    \begin{split}
        F^{\mu\nu}&=E_{(k)}^{[\mu}k^{\nu]}+\epsilon^{\mu\nu\kappa\lambda}k_{\kappa}B^{(k)}_{\lambda} \\
        F^{\ast\mu\nu}&=B_{(k)}^{[\mu}k^{\nu]}+\epsilon^{\mu\nu\kappa\lambda}k_{\kappa}E^{(k)}_{\lambda}
    \end{split}
    \label{eq:Maxwell}
\end{equation}
for
\begin{equation}
    E_{(k)}^{\mu}=E_z\zeta^{\mu}~~~~,~~~~B_{(k)}^{\mu}=B_x\chi^{\mu}+B_y\tau^{\mu}
    \label{eq:field}
\end{equation}
being the electric and magnetic field locally observed in the laboratory. Here $\epsilon^{\mu\nu\kappa\lambda}=-\left(-g\right)^{-1/2}[\mu\nu\kappa\lambda]$, with $[\mu\nu\kappa\lambda]$ the 4D Levi-Civita tensor. Moreover, in the Sweet-Parker configuration, we have:
 \begin{equation}
    \begin{split}
        \text{upstream:~}&B_x\big|_i=b_0,~~B_y\big|_i=0 \\
        \text{downstream:~}&B_x\big|_o=0,~~B_y\big|_o=b_1 \\
        \text{current sheet:~}&B_x\big|_X=B_y\big|_X=0
    \end{split}
    \label{eq:field2}
\end{equation} 
The 4-velocity of plasma, for the Sweet-Parker approach, should be:
\begin{equation}
    u^{\mu}=\gamma\left(k^{\mu}+V_x\chi^{\mu}+V_y\tau^{\mu}\right)
    \label{eq:velocity}
\end{equation}
with
\begin{equation}
    \begin{split}
        \text{upstream:~}&\gamma\big|_i\equiv\gamma_{\rm in}=\left(1-v_{\rm in}^2\right)^{-1/2},~~V_x\big|_i=0,~~V_y\big|_i=-v_{\rm in} \\
        \text{downstream:~}&\gamma\big|_o\equiv\gamma_{\rm in}=\left(1-v_{\rm out}^2\right)^{-1/2},~~V_x\big|_o=v_{\rm out},~~V_y\big|_o=0 \\
        \text{current sheet:~}&\gamma\big|_X=1,~~V_x\big|_X=V_y\big|_X=0 
    \end{split}
    \label{eq:velocity2}
\end{equation}
Here $v_{\rm in/out}$ are the speeds of inflow and outflow plasma while $\gamma_{\rm in/out}$ are their Lorentz factors respectively, observed in the rest frame of laboratory. The GR forms of quasi-stationary condition in Eq.~\eqref{eq:steady} and the substitutions in Eq.~\eqref{eq:L-delta-SR} should be \cite{GRMR1}:
\begin{equation}
    \mathcal{L}_{\hat{k}}q\approx k^{\mu}\partial_{\mu}q\sim 0
    \label{eq:steady_GR}
\end{equation}
and
\begin{equation}
    \mathcal{L}_{\hat{\chi}}q\approx \chi^{\mu}\partial_{\mu}q\sim \frac{q\big|_o-q\big|_X}{L}~~~,~~~
    \mathcal{L}_{\hat{\tau}}q\approx \tau^{\mu}\partial_{\mu}q\sim \frac{q\big|_i-q\big|_X}{L}
    \label{eq:L_delta_GR}
\end{equation}
for $q$ being an arbitrary physical quantity and $\mathcal{L}_{\hat{A}}$ representing the Lee derivatives along the 4-vector $A^{\mu}$. The quasi-stationary condition and substitutions above are different from those applied in Ref.~\cite{CA2017} and Ref.~\cite{CA2018}, which generated different results, as discussed in Ref.~\cite{GRMR1}.

Since we adopt the Sweet-Parker configuration to describe the magnetic reconnection under the generalized MHD scheme, many results in Ref.~\cite{GRMR1}, listed in Appen.~\ref{sec:SP}, are still useful. Because the thermal electromotive force and current inertia are added in Ohm's law and energy-momentum conservation respectively, one should reconsider the third, sixth and seventh steps, similar to the SR case introduced in Sect.~\ref{sec:SR}. In the Sweet-Parker configuration, we are interested in the $\hat{z}$ component of Ohm's law upstream and inside current sheet. Namely, we take care of:
\begin{equation}
    \begin{split}
        {\rm upstream:}&~\zeta_{\mu}F^{\mu\nu}u_{\nu}=0 \\
        {\rm current~sheet:}&~\zeta_{\mu}F^{\mu\nu}u_{\nu}=\frac{1}{\sqrt{-g}}\zeta_{\mu}\partial_{\nu}\left[\frac{\sqrt{-g}\omega}{2{\frak q}^2}u^{(\mu}J^{\nu)}\right]+\frac{\omega}{2{\frak q}^2}\zeta_{\mu}\Gamma^{\mu}_{\kappa\lambda}u^{(\kappa}J^{\lambda)}+\eta\zeta_{\mu}J^{\mu}
    \end{split}
    \label{eq:step3-gen}
\end{equation}
Meanwhile, the $\hat{y}$ and $\hat{x}$ components of energy-momentum conservation should be under consideration, which take the forms of
\begin{equation}
    \frac{1}{\sqrt{-g}}\tau_{\mu}\partial_{\nu}\left(\sqrt{-g}\omega u^{\mu}u^{\nu}+\frac{\sqrt{-g}\omega}{4{\frak q}^2}J^{\mu}J^{\nu}\right)=-\tau_{\mu}\partial^{\mu}p+\tau_{\mu}F^{\mu\nu}J_{\nu}-\omega\tau_{\mu}\Gamma^{\mu}_{\kappa\lambda}\left(u^{\kappa}u^{\lambda}+\frac{1}{4{\frak q}^2}J^{\kappa}J^{\lambda}\right)
    \label{eq:step6-gen}
\end{equation}
and
\begin{equation}
    \frac{1}{\sqrt{-g}}\chi_{\mu}\partial_{\nu}\left(\sqrt{-g}\omega u^{\mu}u^{\nu}+\frac{\sqrt{-g}\omega}{4{\frak q}^2}J^{\mu}J^{\nu}\right)=-\chi_{\mu}\partial^{\mu}p+\chi_{\mu}F^{\mu\nu}J_{\nu}-\omega\chi_{\mu}\Gamma^{\mu}_{\kappa\lambda}\left(u^{\kappa}u^{\lambda}+\frac{1}{4{\frak q}^2}J^{\kappa}J^{\lambda}\right)
    \label{eq:step7-gen}
\end{equation}
respectively. In Sect.~\ref{sec:zamo}, I will show how the thermal electromotive force and current inertia in generalized MHD act on the reconnection rate in GR by discussing one specific example, where some resultant relations between physical properties got from Eq.~\eqref{eq:step1}, \eqref{eq:step2}, \eqref{eq:step4} and \eqref{eq:step5} should be utilized.

\subsection{Observations}
\label{sec:obs}

Here let us consider the magnetic reconnection occurs in some laboratory, in the rest frame of which the reconnection rate is detected to obey the rule shown in Eq.~\eqref{eq:R-1}, neglecting the infinitesimal modifications (which will be discussed in Sect.~\ref{sec:zamo} through a specific example) induced by graviatational effect. Let us try to analyze how the property of reconnection rate is modified if it is detected by an observer who is moving with respect to the laboratory. For the pair plasma under the generalized MHD scheme, only the property of reconnection rate changes, compared to the property under the standard MHD scheme. Thus the modification on the property of outflow speed in observation, which has been discussed in Ref.~\cite{GRMR1} already, is not under consideration in this work. For simplicity, we presume a large Lundquist number ($S\gg 1$), which is satisfied in most astrophysical systems \cite{yamada2009}, such that the reconnection rate detected in the laboratory obeys:
\begin{equation}
    R\equiv \frac{v_{\rm in}}{v_{\rm A}} \simeq \sqrt{\frac{\omega\sqrt{1+\sigma_0}}{4{\frak q}^2L^2}} \simeq \left\{
    \begin{split}
        \frac{\omega^{1/2}}{2{\frak q}L}~~~~~~~&\text{LM} \\
        \frac{\left(\omega^2\sigma_0\right)^{1/4}}{2{\frak q}L}~~~&\text{HM}
    \end{split} \right.
    \label{eq:R-1-1}
\end{equation}
where "LM" and "HM" denote the limits of low magnetization ($\sigma_0\ll 1$) and high magnetization ($\sigma_0\gg 1$) respectively. In Eq.~\eqref{eq:R-1-1}, the inflow speed, the length of current sheet, the local magnetization and the Alfv{\'e}n velocity depend on the relation between $k^{\mu}$ and $n^{\mu}$. Based on the discussions in Ref.~\cite{GRMR1}, we have:
\begin{itemize}
\item Observed inflow speed:
\begin{equation}
    v_{\rm in,obs}=\left|-\frac{u^{\mu}\big|_{i/o}+n_{\nu}u^{\nu}\big|_{i/o} n^{\mu}}{n_{\nu}u^{\nu}\big|_{i/o}}+\frac{k^{\mu}+n_{\nu}k^{\nu}n^{\mu}}{n_{\nu}k^{\nu}}\right|
    \label{eq:v_obs}
\end{equation}

\item Observed length of current sheet:
\begin{equation}
    L_{\rm obs}=\frac{L}{\left|\chi^{\mu}+n_{\nu}\chi^{\nu}n^{\mu}\right|}
    \label{eq:L_obs}
\end{equation}

\item Observed magnetization and Alfv{\'e}n velocity:
\begin{equation}
    \sigma_{\rm 0,obs}=\frac{\left(-n_{\nu}F^{\ast j\nu}\big|_i\right)^2}{\omega}~~~,~~~v_{\rm A,obs}=\sqrt{\frac{\sigma_{\rm 0,obs}}{1+\sigma_{\rm 0,obs}}}
    \label{eq:vA_obs}
\end{equation}
\end{itemize}
We would like to determine the factor of modification:
\begin{equation}
    \mathscr{F}_{R}\equiv \frac{R_{\rm obs}}{\sqrt{\frac{\omega\sqrt{1+\sigma_{\rm 0,obs}}}{4{\frak q}^2L_{\rm obs}^2}}} \simeq \left\{
    \begin{split}
        \frac{v_{\rm in,obs}}{v_{\rm A,obs}}\Bigg/ \frac{\omega^{1/2}}{2{\frak q}L_{\rm obs}}~~~~~~~&\text{LM} \\
        \frac{v_{\rm in,obs}}{v_{\rm A,obs}}\Bigg/ \frac{\left(\omega^2\sigma_{\rm 0,obs}\right)^{1/4}}{2{\frak q}L_{\rm obs}}~~~&\text{HM}
    \end{split} \right.
    \label{eq:rela}
\end{equation}
which equals 1 when the magnetic reconnection occurs just in the rest frame of observer ($n^{\mu}=k^{\mu}$). Specific examples about $\mathscr{F}_{R}$ will be shown in Sect.~\ref{sec:exm_obs}.

\section{GR description: examples}
\label{sec:example}

\subsection{Metric and rest frames}
\label{sec:metric}

The magnetic reconnection process is considered to occur in a stationary, axis-symmetric spacetime, in which equations could be written compactly in 3+1 formalism \cite{MacDonald:1982zz}. The line element in 3+1 formalism takes the form of:
\begin{equation}
    ds^2=-\alpha^2dt^2+\sum_{i=1}^{3}h_{i}^2\left(dx^i-\omega^{i}dt\right)^2
    \label{eq:element}
\end{equation}
with $\alpha$ the lapse function, $h_i$ the scale factors of the coordinates $x^i$ and $\omega^i$ the velocity corresponding to a frame dragging. In this work, magnetic reconnection is assumed to happen near a rotating black hole described by Kerr metric in Boyer-Lindquist (BL) coordinates $(t,r,\theta,\phi)$, for which:
\begin{equation}
    \alpha=\sqrt{\frac{\Delta\Sigma}{A}},~~h_r=\sqrt{\frac{\Sigma}{\Delta}},~~h_{\theta}=\sqrt{\Sigma},~~h_{\phi}=\sqrt{\frac{A}{\Sigma}}\sin\theta,~~
    \omega^r=\omega^{\theta}=0,~~\omega^{\phi}=\frac{2ar}{A},
\end{equation} 
where $\Delta=r^2-2r+a^2$, $\Sigma=r^2+a^2\cos^2\theta$ and $A=\left(r^2+a^2\right)^2-\Delta a^2\sin^2\theta$. The determinant of metric tensor obeys: $\sqrt{-g}=\alpha h_1 h_2 h_3=\Sigma\sin\theta$. 

The rest frame of ZAMOs could be defined via the normal tetrad taking the form of:
\begin{equation}
    \hat{e}_{(t)}^{\mu}=\frac{1}{\alpha}\left(\partial_t^{\mu}+\omega^{\phi}\partial_{\phi}^{\mu}\right);~~
    \hat{e}_{(r)}^{\mu}=\frac{1}{h_r}\partial_{r}^{\mu},~~
    \hat{e}_{(\theta)}^{\mu}=\frac{1}{h_{\theta}}\partial_{\theta}^{\mu},~~
    \hat{e}_{(\phi)}^{\mu}=\frac{1}{h_{\phi}}\partial_{\phi}^{\mu}
    \label{eq:ZAMO}
\end{equation}
While it would be more important to consider the rest frame of fluid moving on the equatorial, which could be defined via the normal tetrad as \cite{Work0,Work1}:
\begin{equation}
    \begin{split}
        e_{[0]}^{\mu}&=\hat{\gamma}_s \left[\hat{e}_{(t)}^{\mu}+\hat{v}_s^{(r)}\hat{e}_{(r)}^{\mu}+\hat{v}_s^{(\phi)}\hat{e}_{(\phi)}^{\mu}\right]; \\
        e_{[1]}^{\mu}&=\frac{1}{\hat{v}_s}\left[\hat{v}_s^{(\phi)}\hat{e}_{(r)}^{\mu}-\hat{v}_s^{(r)}\hat{e}_{(\phi)}^{\mu}\right],~~
        e_{[2]}^{\mu}=\hat{e}_{(\theta)}^{\mu}, \\
        e_{[3]}^{\mu}&=\hat{\gamma}_s\left[\hat{v}_s\hat{e}_{(t)}^{\mu}+\frac{\hat{v}_s^{(r)}}{\hat{v}_s}\hat{e}_{(r)}^{\mu}+\frac{\hat{v}_s^{(\phi)}}{\hat{v}_s}\hat{e}_{(\phi)}^{\mu}\right]
    \end{split}
    \label{eq:plasma}
\end{equation}
where $\hat{v}_s=\sqrt{\left(\hat{v}_s^{(r)}\right)^2+\left(\hat{v}_s^{(\phi)}\right)^2}$ is the speed with $\hat{v}_s^{(r)}$ and $\hat{v}_s^{(r)}$ being the components of 3-velocity in the view of ZAMOs and $\hat{\gamma}_s$ is the Lorentz factor. It is easy to figure out that $e_{[3]}^{\mu}$ and $e_{[1]}^{\mu}$ are parallel and perpendicular to the moving direction of fluid respectively.

\subsection{ZAMOs laboratory with azimuthal current sheet}
\label{sec:zamo}

Align with the discussions in Ref.~\cite{CA2017,CA2018,GRMR1}, let us focus on the magnetic reconnection occuring in the rest frame of ZAMOs (refered to as ZAMOs laboratory henceforth) in this section, with $\hat{x}$ and $\hat{y}$ axes in the Sweet-Parker configuration posited azimuthally and radially in the BL coordinates respectively. That is to say, we presume:
\begin{equation}
    k^{\mu}=\hat{e}_{(t)}^{\mu};~~\chi^{\mu}=\hat{e}_{(\phi)}^{\mu},~~\tau^{\mu}=\hat{e}_{(r)}^{\mu},~~\zeta^{\mu}=\hat{e}_{(\theta)}^{\mu}
    \label{eq:tetrad_z}
\end{equation} 
Subsequently, the specific forms of quasi-stationary condition and the substitutions are:
\begin{equation}
    \mathcal{L}_{\hat{k}}q\approx\left(\frac{1}{\alpha}\partial_t+\frac{\omega^{\phi}}{\alpha}\partial_{\phi}\right)q \sim 0
    \label{eq:steady_z}
\end{equation}
and
\begin{equation}
    \mathcal{L}_{\hat{\chi}}q\approx\frac{1}{h_{\phi}}\partial_{\phi}q\sim \frac{q\big|_o-q\big|_X}{L}~~,~~
    \mathcal{L}_{\hat{\tau}}q\approx\frac{1}{h_r}\partial_{r}q\sim \frac{q\big|_i-q\big|_X}{\delta}
    \label{eq:L_delta_z}
\end{equation}
where $q$ is an arbitrary physical quantity. Eq.~\eqref{eq:L_delta_z} is different from the substitutions adopted in Ref.~\cite{CA2017,CA2018}, which basically argued that:
\begin{equation}
    \frac{1}{r}\partial_{\phi}q\sim\frac{q\big|_o-q\big|_X}{L}~~,~~
    \partial_{r}q\sim\frac{q\big|_i-q\big|_X}{\delta}
    \label{eq:L_delta_CA}
\end{equation}
In Ref.~\cite{GRMR1}, one can find the differences between the resultant expressions in Sweet-Parker model by applying Eq.~\eqref{eq:L_delta_z} or Eq.~\eqref{eq:L_delta_CA}. In the following calculations we always adopt Eq.~\eqref{eq:steady_z} and Eq.~\eqref{eq:L_delta_z} unless otherwise specified. 

The $\hat{z}$ component of Ohm's law upstream and inside the current sheet gives:
\begin{equation}
    \begin{split}
        \text{upstream:}&~\gamma_{\rm in}\left(e_z+v_{\rm in}b_0\right)=0 \\
        \text{current sheet:}&~J_z=\frac{E_z}{\eta+\Lambda}\simeq \frac{e_z}{\eta+\Lambda}\equiv j_z
    \end{split}
    \label{eq:step3_z-gen}
\end{equation}
where $e_z$ is constant approximately over the local scale (see Eq.~\eqref{eq:step1-res}). The term $\frac{\omega}{2{\frak q}^2}\zeta_{\mu}\Gamma^{\mu}_{\kappa\lambda}u^{(\kappa}J^{\lambda)}$ vanishes in Eq.~\eqref{eq:step3_z-gen} approximately because $V_y\ll 1$ (see Eq.~\eqref{eq:step5-res}). The thermal-inertia parameter $\Lambda$ in GR reads: 
\begin{equation}
    \Lambda=\frac{\omega}{4{\frak q}^2}\left[\left(\frac{1}{\alpha}\partial_t+\frac{\omega^{\phi}}{\alpha}\partial_{\phi}\right)\gamma+\frac{1}{h_{\phi}}\partial_{\phi}\gamma V_x\right]\simeq \frac{\omega\gamma_{\rm out}v_{\rm out}}{4{\frak q}^2L}
    \label{eq:Lambda_z}
\end{equation}
The final expression is identical to its SR form shown in Eq.~\eqref{eq:Lambda-1}, for which the quasi-stationary condition in Eq.~\eqref{eq:steady_z} and substitutions in Eq.~\eqref{eq:L_delta_z} are applied. If one adopts Eq.~\eqref{eq:L_delta_CA} for substitutions, the resultant expression of thermal-inertial parameter would become
\begin{equation}
    \Lambda \simeq \frac{\omega\gamma_{\rm out}v_{\rm out}}{4{\frak q}^2L}\frac{r}{h_{\phi}}
\end{equation}
It is identical to the Eq.~(44) in Ref.~\cite{CA2018}, where $r/h_{\phi}$ was refered to as the factor of modification induced by spacetime curvature. However, according to the opinion in Ref.~\cite{GRMR1}, no effect of spacetime curvature should exist under the scheme of standard MHD in GR. Situations would be the same under the generalized MHD scheme. 

Since the current flows along $\zeta^{\mu}$ purely in the current sheet for the Sweet-Parker approach (conclusion of Amp{\`e}re's law in Eq.~\eqref{eq:step4} as well), we generally have:
\begin{equation}
    \tau_{\mu}J^{\mu}J^{\nu}=\chi_{\mu}J^{\mu}J^{\nu}=0
    \label{eq:0}
\end{equation}
Subsequently, the $\hat{y}$ component of energy-momentum conservation becomes:
\begin{equation}
    \frac{1}{h_r}\partial_rp + \omega\mathcal{O}_1\left(\Gamma\right) + \frac{\omega j_z^2}{4{\frak q}^2}\mathcal{O}_2\left(\Gamma\right)=J_zB_x
    \label{eq:step6_z-gen}
\end{equation}
with
\begin{equation}
    \mathcal{O}_1\left(\Gamma\right)\simeq \frac{1}{h_r^3}\partial_r \ln\alpha~~,~~
    \mathcal{O}_2\left(\Gamma\right)\simeq -\frac{1}{h_r}\partial_r \ln h_{\theta}
\end{equation}
The quasi-stationary condition in Eq.~\eqref{eq:steady_z} is applied in Eq.~\eqref{eq:step6_z-gen}. Utilizing the relation between $J_z$ and $B_x$ shown in Eq.~\eqref{eq:step4-res} and the substitutions in Eq.~\eqref{eq:L_delta_z}, one gets:
\begin{equation}
    \frac{p_X}{b_0^2/2}\simeq \frac{\left(\alpha^2h_{\theta}^2h_{\phi}^2\right)\bigg|_i}{\left(\alpha^2h_{\theta}^2h_{\phi}^2\right)\bigg|_{\varepsilon_2}}-
    \frac{2}{\sigma_0}\frac{\ln\alpha\big|_i-\ln\alpha\big|_X}{h_r^2\big|_{\varepsilon_3}}+ \frac{j_z^2}{2{\frak q}^2\sigma_0}\left(\ln h_{\theta}\big|_i-\ln h_{\theta}\big|_X\right) \simeq 1
    \label{eq:step6_z-gen-res}
\end{equation}
where $\varepsilon_2$ and $\varepsilon_3$ are two different points between $i$ and $X$ (see Eq.~\eqref{eq:mean}). Resultantly, Eq.~\eqref{eq:step6_z-gen} gives $p_X\simeq b_0^2/2$, the so called incompressibility of current sheet shown in Eq.~\eqref{eq:step6-res} as well, when $i$ and $X$ are extremely close (namely $\delta\ll 1$). What is more, one can see that the current inertia, which disappears in the SR case, induces perturbation (though infinitesimal) in the gravitational field. In SR, since current flows along $\hat{z}$ axis only, the current inertia, acting just like the "inertial mass" of current, has no effect on the equations in $\hat{x}$ and $\hat{y}$ directions. However, in GR, the current inertia is involved with the affine connection, acting in analogous to the "gravitational mass" of current. Hence, different from the SR case, we could not simply discard the current inertia in the Sweet-Parker configuration when handling the generalized MHD equations in GR.

Eq.~\eqref{eq:step7-gen}, namely the $\hat{x}$ component of energy-momentum conservation, gives:
\begin{equation}
    \frac{1}{h_{\phi}}\partial_{\phi}\left(\omega\gamma^2V_x^2+p\right)\simeq -J_zB_y
    \label{eq:step7_z-gen}
\end{equation}
after applying $V_y\ll 1$ and the quasi-stationary condition. Terms $\chi_{\mu}\Gamma^{\mu}_{\kappa\lambda}u^{\kappa}u^{\lambda}$ and $\chi_{\mu}\Gamma^{\mu}_{\kappa\lambda}J^{\kappa}J^{\lambda}$ both vanish here. It should be noticed here that the relation between $p_X$ and $b_0$ in Eq.~\eqref{eq:step6_z-gen-res} are different from the relation given by the standard MHD equations (see Eq.~\eqref{eq:step6-res}). One can not just copy Eq.~\eqref{eq:step7-res} and Eq.~\eqref{eq:D_SP} directly although Eq.~\eqref{eq:step7_z-gen} is identical to the resultant expression of Eq.~\eqref{eq:step7}. Utilizing Eq.~\eqref{eq:step6_z-gen-res}, Eq.~\eqref{eq:step4-res} and the substitutions in Eq.~\eqref{eq:L_delta_z}, one gets:
\begin{equation}
    v_{\rm out}\simeq \sqrt{\frac{\mathscr{D}\sigma_0}{1+\mathscr{D}\sigma_0}}
\end{equation}
with
\begin{equation}
    \mathscr{D}\equiv \frac{1}{2}\left[\frac{\left(\alpha^2h_{\theta}^2h_{\phi}^2\right)\bigg|_i}{\left(\alpha^2h_{\theta}^2h_{\phi}^2\right)\bigg|_{\varepsilon_2}}-
    \frac{2}{\sigma_0}\frac{\ln\alpha\big|_X-\ln\alpha\big|_i}{h_r^2\big|_{\varepsilon_3}} + \frac{j_z^2}{2{\frak q}^2\sigma_0}\left(\ln h_{\theta}\big|_i-\ln h_{\theta}\big|_X\right) +
    \frac{\left(\alpha h_{\theta}h_{\phi}\right)\big|_o}{\left(\alpha h_{\theta}h_{\phi}\right)\big|_{\varepsilon_1}}\right] \simeq 1
    \label{eq:D}
\end{equation}
for $\varepsilon_1$ being some point between $i$ and $X$ (see Eq.~\eqref{eq:mean} and Eq.~\eqref{eq:step4-res}). When considering the low magnetization limit ($\sigma_0\ll 1$, hence $\gamma_{\rm out}\simeq 1$ and $\gamma_{\rm in}\simeq 1$), the reconnection rate becomes:
\begin{equation}
    R\equiv \frac{v_{\rm in}}{v_{\rm A}}\simeq \mathscr{CD}\frac{\delta}{L}\simeq \mathscr{C}^{-1}\frac{\eta+\Lambda}{\delta v_{\rm A}} \simeq 
    \mathscr{D}^{1/2}\left(\frac{1}{S}+\frac{\Lambda}{Lv_{\rm A}}\right)^{1/2} \simeq \frac{\mathscr{D}^{3/4}\omega^{1/2}}{2{\frak q}L}
    \label{eq:step7_z-gen-res}
\end{equation}
A large lundquist number is adopted for the final expression. These results are identical to Eq.~\eqref{eq:vout-SR} and Eq.~\eqref{eq:R-1} when $o$ and $X$ are extremely close (namely $L\ll 1$), in which case $\mathscr{D}$ approaches 1. It is consistent with the opinion reiterated in Ref.~\cite{GRMR1} that the spacetime curvature, corresponding to the higher order derivatives and non-linear terms of $g_{\mu\nu}$, has no effect on the magnetic reconnection process. While the modifications of the properties induced by gravitational effect tend to be infinitesimal when the magnetic reconnection occurs in a local scale.

\subsection{Examples of observations}
\label{sec:exm_obs}

In this section, I would like to show two examples about the modification on the property of reconnection rate in the case that the reconnection rate is detected by an observer who is moving with respect to the laboratory. As discussed in Sect.~\ref{sec:zamo}, modifications induced by gravitational effect are negligible whenever $\delta\ll L\ll 1$. However, the property of reconnection rate would be modified significantly if it is not detected in the rest frame of laboratory and this modification should be non-vanishing. Based on the discussions in Sect.~\ref{sec:obs}, reconnection rate should obey Eq.~\eqref{eq:R-1} if it is detected in the rest frame of laboratory when a large lundquist number is presumed. Since the property of outflow speed given by generalized MHD equations is the same as that given by standard MHD equations, the modified property of outflow speed, which has been discussed in Ref~\cite{GRMR1} already, is not under consideration in this work.

First, let us consider the magnetic reconnection occurring in the rest frame of plasma moving on the equatorial plane (called plasma laboratory henceforth) and it is observed by ZAMOs. That is to say, we choose:
\begin{equation}
    k^{\mu}=e_{[0]}^{\mu}~~~,~~~n^{\mu}=\hat{e}_{(t)}^{\mu}
\end{equation}
The current sheet is posited in an arbitrary direction on the equatorial plane such that:
\begin{equation}
    \chi^{\mu}=\cos\xi_B e_{[3]}^{\mu}+\sin\xi_B e_{[1]}^{\mu}~~,~~
    \tau^{\mu}=e_{[2]}^{\mu}~~,~~
    \zeta^{\mu}=\cos\xi_B e_{[1]}^{\mu}-\sin\xi_B e_{[3]}^{\mu}
\end{equation}
The orientation angle $\xi_B$ is defined to be the angle between $\chi^{\mu}$ and $e_{[3]}^{\mu}$. In other words, it is the angle between the direction of current sheet and the moving direction of plasma \cite{Work1}. One could calculate the inflow speed, length of current sheet, local magnetization and Alfven velocity observed by ZAMOs based on Eq.~\eqref{eq:v_obs}--\eqref{eq:vA_obs}. The results are:
\begin{equation}
    \begin{split}
        v_{\rm in,obs}&=\hat{\gamma}_s^{-1} v_{\rm in} \\
        ~~~~~~~~~ \\
        L_{\rm obs}&=\frac{L}{\left(\hat{\gamma}_s^2\cos^2\xi_B+\sin^2\xi_B\right)^{1/2}} \\
        ~~~~~~~~~~~~\\
        \sigma_{\rm 0,obs}&\simeq \left(\cos^2\xi_B+\hat{\gamma}_s^2\sin^2\xi_B\right)\sigma_0 \\
        ~~~~~~\\
        v_{\rm A,obs}&\simeq 
        \begin{cases}
            ~\left(\cos^2\xi_B+\hat{\gamma}_s^2\sin^2\xi_B\right)^{1/2}v_{\rm A}~~~&\text{LM} \\
            ~v_{\rm A}~~~~~~~~~~~~~~~~~~~~~~~~~~~~~~~~~~~&\text{HM}
        \end{cases}
    \end{split}
    \label{eq:plasma_zamo}
\end{equation}
Applying Eq.~\eqref{eq:plasma_zamo} to Eq.~\eqref{eq:rela}, we get the factor of modification:
\begin{equation}
    \mathscr{F}_{\rm obs}\simeq \left\{
    \begin{split}
        \hat{\gamma}_s^{-3}\left(1-\hat{v}_s^{2}\cos^2\xi_B\right)^{-1/2}\left(1-\hat{v}_s^{2}\sin^2\xi_B\right)^{-1/2}~~~~&\text{LM} \\
        \hat{\gamma}_s^{-5/2}\left(1-\hat{v}_s^{2}\cos^2\xi_B\right)^{-1/4}\left(1-\hat{v}_s^{2}\sin^2\xi_B\right)^{-1/2}~~~~&\text{HM}
    \end{split} \right.
\end{equation}
One can also consider the factor of modification in the case of $\xi_B=0$ such that the current sheet is parallel to the moving direction of plasma (referred to as para case) or in the case of $\xi_B=\pi/2$ such that the current sheet is perpendicular to the moving direction of plasma (referred to as perp case). The results are:
\begin{equation}
    \mathscr{F}_{\rm obs}\simeq 
    \begin{cases}
        ~\hat{\gamma}_s^{-3/2}~~~&\text{HM in perp case} \\
        ~\hat{\gamma}_s^{-2}~~~&\text{otherwise}
    \end{cases} 
\end{equation}
Because $\hat{\gamma}_s\geq 1$ is always satisfied, the reconnection rates observed by ZAMOs are always lower than their expectations when the magnetic reconnection occurs in the rest frame of plasma laboratory.

Next, let us consider the magnetic reconnection occurring in ZAMOs laboratory while the reconnection rate is detected by a static observer in BL coordinates. That is to say, we choose:
\begin{equation}
    k^{\mu}=\hat{e}_{(t)}^{\mu}~~~,~~~n^{\mu}=\left[\alpha^2-\left(h_{\phi}\omega^{\phi}\right)^2\right]^{-1/2}\partial_t^{\mu}
\end{equation}
The observer $n^{\mu}$ is well defined outside the ergosphere only. The current sheet is posited based on Eq.~\eqref{eq:tetrad_z}. Untilizing Eq.~\eqref{eq:v_obs}--\eqref{eq:vA_obs} again, we get:
\begin{equation}
    \begin{split}
        v_{\rm in,obs}&=\alpha^{-1}\left[\alpha^2-\left(h_{\phi}\omega^{\phi}\right)^2\right]^{1/2}v_{\rm in} \\
        ~~~~~~~~~ \\
        L_{\rm obs}&=\alpha^{-1}\left[\alpha^2-\left(h_{\phi}\omega^{\phi}\right)^2\right]^{1/2}L \\
        ~~~~~~~~~~~~\\
        \sigma_{\rm 0,obs}&\simeq \alpha^{-2}\left[\alpha^2-\left(h_{\phi}\omega^{\phi}\right)^2\right]\sigma_0 \\
        ~~~~~~\\
        v_{\rm A,obs}&\simeq 
        \begin{cases}
            ~\alpha^{-1}\left[\alpha^2-\left(h_{\phi}\omega^{\phi}\right)^2\right]^{1/2}v_{\rm A}~~~&\text{LM} \\
            ~v_{\rm A}~~~~~~~~~~~~~~~~~~~~~~~~~~~~~~~~~&\text{HM}
        \end{cases}
    \end{split}
    \label{eq:zamo_static}
\end{equation}
In this case, the factor of modification should be:
\begin{equation}
    \mathscr{F}_{\rm obs}\simeq \left\{
    \begin{split}
        \alpha^{-1}\left[\alpha^2-\left(h_{\phi}\omega^{\phi}\right)^2\right]^{1/2}\simeq 1-2a^2 r^{-4}\sin^2\theta -4a^2 r^{-5}\sin^2\theta +...~~~~&\text{LM} \\
        \alpha^{-3/2}\left[\alpha^2-\left(h_{\phi}\omega^{\phi}\right)^2\right]^{3/4}\simeq 1-3a^2 r^{-4}\sin^2\theta-6a^2 r^{-5}\sin^2\theta+...~~~~&\text{HM}
    \end{split} \right.
    \label{eq:zamo_static-res}
\end{equation}
Recall that $n^{\mu}$ is well defined outside the ergosphere only. Thus the factor of modification in Eq.~\eqref{eq:zamo_static-res} is always positive. The static observer would never get a negative reconnection rate which is counterintuitive. Here is the story narrated by Eq.~\eqref{eq:zamo_static-res}: when static observers near a Kerr black hole detect the reconnection rates of magnetic reconnection processes occurring in the rest frame of ZAMOs, the values of reconnection rate they get would always be lower than the values they predict based on the reconnection model introduced in Sect.~\ref{sec:SR}.

\section{Summary}
\label{sec:sum}

In this work, I extend the discussions in Ref.~\cite{GRMR1} to the generalized MHD scheme, in order to describe the fast magnetic reconnection induced by thermal electromotive force with Sweet-Parker configuration in general relativity. As a sequel, the notations, basic setups and methods introduced in Ref.~\cite{GRMR1} are inherited. 

I revisit the reconnection model in SR by analyzing the generalized MHD equations for pair plasma with the Sweet-Parker configuration adopted. How the two added terms, the thermal electromotive force and current inertia, act in the process and how the thermal-inertial parameter, corresponding to the thermal electromotive force, increases the reconnection rate are exhibited. I list the equations of generalized MHD in GR. By discussing the example of ZAMOs laboratory with the current sheet posited azimuthally, I try to calculate the propertes of reconnection rate and outflow speed which are modified by gravitational effect. The results indicate that the modifications on properties induced by gravitational field tend to be infinitesimal when the local scale, within which the magnetic reconnection occurs, is tiny, which is consistent with the results in Ref.~\cite{GRMR1}. What is more, I figure out that the current inertia, which has no effect in SR because it displays like the "inertial mass" of current, could not be discarded simply in GR, because it is involved with the affine connection, acting in analogous to the "gravitational mass" of current. Finally, I calculate the factor of modification when the reconnection rate is detected not in the rest frame of laboratory. Two specific examples, the plasma laboratory observed by ZAMOs and the ZAMOs laboratory observed by static observer, are discussed. Consistent with the conclusion in Ref.~\cite{GRMR1}, modifications induced by observation are significant and should not be neglected. While the observer would never get a negative reconnection rate from detection which is counterintuitive.

As exploratory studies, this work and Ref.~\cite{GRMR1} opposite the conclusions in Ref.~\cite{CA2017,CA2018,Fan:2024fcy,Fan:2024rsa} which described the magnetic reconnection in Sweet-Parker configuration differently in GR. Moreover, the choices of quasi-stationary condition and the approximations which substitute derivatives by finite differences in these two groups of works are different. One cannot affirm whose opinion is correct subjectively. All in all, further analyses would be necessary to compare these two groups of works or even propose a new description before getting the best answer.

Moreover, the magnetic reconnection described in the Sweet-Parker configuration is oversimplified. For example, it is ideal to set that the magnetic strength vanishes strictly within current sheet. Also, current would never become completely null outside current sheet. Magnetic reconnection process occurs in a real astrophysical system would be much more complicated. It is predictable that more analytical models of magnetic reconnection would be established and discussed under the GR scheme in order to get closer to the processes occurring in accretion systems around real astrophysical black holes.

In reality, the effect of observation discussed in this work and Ref.~\cite{GRMR1} could not be verified yet from observation in astronomy, although the frequent occurrences of magnetic reconnection in accretion flow near black holes have been widely accepted and supported by numerical simulations already \cite{Jia:2023iup,Davelaar:2023dhl,Aimar2023,Ripperda2020,Yuan2024-1,Yuan2024-2}. With the development of observational precision, it is worth expecting that the magnetic reconnection processes occurring near black holes could be observed and the properties of outflow speed and reconnection rate could be detected. Mysteries of either the magnetic reconnection in astrophysical scenarios or the gravitational effect could be revealed by then.

\section*{Acknowledgement}

I appreciate the support and constructive suggestions from Prof.~Bin Chen. I would like to thank Dr.~Yuedan Wang for her encouragement.

\appendix

\section{Equations in generalized MHD}
\label{sec:gMHD}

The equations in generalized MHD in one-fluid forms were derived from two-fluids forms in Ref.~\cite{Koide2009}. Under the scheme of generalized MHD, positively and negatively charged particles are treated as perfect fluid separately. Except for the Maxwell's equations, the mass conservation and equations of motion (or energy-momentum conservation equivalently) for two fluids should be considered\footnote{The equations are written in covariant forms as expressed in Ref.~\cite{Koide2009}}:
\begin{equation}
    \begin{aligned}
        \textbf{mass conservation:}&~\partial_{\nu}\left(n_{\pm}u_{\pm}^{\nu}\right)=0 \\
        \textbf{equation of motion:}&~\partial_{\nu}\left(\omega_{\pm}u_{\pm}^{\mu}u_{\pm}^{\nu}\right)+\partial^{\mu}p_{\pm}=\pm en_{\pm}\eta^{\mu\kappa}u_{\pm}^{\nu}F_{\kappa\nu}\pm R^{\mu}
    \end{aligned}
    \label{eq:gMHD1}
\end{equation}
Here, $R^{\mu}$ represents the frictional 4-force density which causes the energy-momentum exchange between two fluids. While $\eta^{\mu\kappa}$ is the metric tensor in flat spacetime. Since the positively and negatively charged particles are considered separately in generalized MHD, Ohm's law is actually included in and will be derived from the equations of motion.

It is more convenient to describe the generalized MHD in one-fluid form, from which the Hall effect, current inertia, thermal electromotive force and the thermal exchange could be apparently seen. For this purpose, we should define the average variables as follows:
\begin{equation}
    \begin{aligned}
        \text{average mass density:}&~~\rho=m_+n_++m_-n_- \\
        ~~~~~~~~~~~~~~~~~~~~\\
        \text{average number density:}&~~n=\frac{m_+n_++m_-n_-}{m_++m_-} \\
        ~~~~~~~~~~~~~~~~~~~~\\
        \text{average 4-velocity:}&~~u^{\mu}=\frac{m_+n_+u_+^{\mu}+m_-n_-u_-^{\mu}}{m_+n_++m_-n_-} \\
        ~~~~~~~~~~~~~~~~~~~~\\
        \text{current density:}&~~J^{\mu}=e\left(n_+u_+^{\mu}-n_-u_-^{\mu}\right)
    \end{aligned}
    \label{ave_variable}
\end{equation}
Besides, the average and difference variables with respect to the enthalpy density should be defined as follows:
\begin{equation}
    \begin{aligned}
        \omega &= n^2\left(\frac{\omega_+}{n_+^2}+\frac{\omega_-}{n_-^2}\right)~~~,~~~
        \omega^{\dagger} = \frac{n^2}{m}\left(\frac{2m_-\omega_+}{n_+^2}-\frac{2m_+\omega_-}{n_-^2}\right) \\
        ~~~~~~~~~~~~~~~~~~~~\\
        \omega^{\ddagger} &= \frac{n^2}{m^2}\left(\frac{4m_-^2\omega_+}{n_+^2}+\frac{4m_+^2\omega_-}{n_-^2}\right)~~~,~~~ 
        \omega^{\sharp} = \frac{n^2}{m^3}\left(\frac{8m_-^3\omega_+}{n_+^2}-\frac{8m_+^3\omega_-}{n_-^2}\right)
    \end{aligned}
    \label{ave_enthalpy}
\end{equation}
Here $m=m_++m_-$. The $\omega$ in Eq.~\eqref{ave_enthalpy} is just the average enthalpy we generally choose in standard MHD. The average and difference mass ratio are defined to be:
\begin{equation}
    \mu=\frac{m_+m_-}{m^2}~~~,~~~\mu^{\dagger}=\frac{m_+-m_-}{m}
    \label{ave_mass}
\end{equation}
with which one can find out some useful relations below:
\begin{equation}
    \omega^{\ddagger}=4\mu\omega-\mu^{\dagger}\omega^{\dagger}~~~,~~~\omega^{\sharp}=-8\mu\mu^{\dagger}\omega+4(1-3\mu)\omega^{\dagger}
    \label{eq:ome_mu}
\end{equation}
Using the average and difference variables defined in Eq.~\eqref{ave_variable}--\eqref{ave_mass} and also applying the relations in Eq.~\eqref{eq:ome_mu}, the equations in Eq.~\eqref{eq:gMHD1} could then be written in one-fluid form. The mass conservation now becomes:
\begin{equation}
    \partial_{\mu}\left(\rho u^{\mu}\right)=0
\end{equation}
which is just identical to the equation of mass conservation in the standard MHD. While the equations of motion become:
\begin{equation}
    \partial_{\nu}\left[\omega u^{\mu}u^{\nu}+\frac{\omega^{\ddagger}}{4\frak{q}^2}J^{\mu}J^{\nu}+\frac{\omega^{\dagger}}{\frak{q}}u^{(\mu}J^{\nu)}\right]+\partial^{\mu}p=F^{\mu\nu}J_{\nu}
   \label{eq:g_mom-con}
\end{equation}
and
\begin{equation}
    \frac{1}{2\frak{q}}\partial_{\nu}\left[\frac{\omega^{\ddagger}}{\frak{q}}u^{(\mu}J^{\nu)}+\omega^{\dagger}u^{\mu}u^{\nu}+\frac{\omega^{\sharp}}{4\frak{q}^2}J^{\mu}J^{\nu}\right]=\frac{1}{2\frak{q}}\partial^{\mu}\left(\mu^{\dagger}p-\Delta p\right)+F^{\mu\nu}u_{\nu}-\frac{\mu^{\dagger}}{\frak{q}}F^{\mu\nu}J_{\nu}+\frac{R^{\mu}}{\frak{q}}
    \label{eq:g_Ohm}
\end{equation}
where $\frak{q}=en$ is the latent charge while $\Delta p=p_+-p_-$ is the difference of pressure. Eq.~\eqref{eq:g_Ohm} is just the generalized Ohm's law, where the friction could be generally written in the form of:
\begin{equation}
    R^{\mu}=-\frak{q}\eta\left[J^{\mu}-(1+\Theta)J_{\nu}u^{\nu}u^{\mu}\right]
    \label{eq:fric}
\end{equation}
for $\eta$ being the electrical resistivity and $\Theta$ being the thermal exchange rate. From Eq.~\eqref{eq:g_mom-con} and \eqref{eq:g_Ohm}, one can see that the Hall effect (the term proportional to $\frac{1}{\frak{q}}F^{\mu\nu}J_{\nu}$), current inertia (the terms proportional to $J^{\mu}J^{\nu}$) and thermal electromotive force (the terms proportional to $u^{(\mu}J^{\nu)}$) are added by multiplying the difference variables. What people generally call the standard MHD is just a special case that the friction is dorminant in the equations of motion.

Now let us simplify the equations under the limit of pair plasma and ion-electron plasma. A widely applied assumption is:
\begin{equation}
    \omega^{\dagger} = \frac{n^2}{m}\left(\frac{2m_-\omega_+}{n_+^2}-\frac{2m_+\omega_-}{n_-^2}\right) \approx 0
    \label{eq:assump}
\end{equation}
with which we have:
\begin{equation}
    \omega^{\ddagger}\approx 4\mu\omega~~~,~~~\omega^{\sharp}\approx -8\mu\mu^{\dagger}\omega
    \label{eq:ome_mu2}
\end{equation}
In a nearly neutral plasma ($n_+\approx n_-$), the assumption in Eq.~\eqref{eq:assump} means the enthalpies of two fluids are mainly composed of rest masses. For the pair plasma where $m_+\approx m_-$ such that $\mu\approx 1/4$ and $\mu^{\dagger}\approx 0$, the equations of motion become:
\begin{equation}
    \partial_{\nu}\left(\omega u^{\mu}u^{\nu}+\frac{\omega}{4\frak{q}}J^{\mu}J^{\nu}\right)+\partial^{\mu}p=F^{\mu\nu}J_{\nu}
\end{equation} 
and
\begin{equation}
    F^{\mu\nu}u_{\nu}-\frac{1}{2\frak{q}}\partial^{\mu}\Delta p=\frac{1}{2\frak{q}}\partial_{\nu}\left[\frac{\omega}{\frak{q}}u^{(\mu}J^{\nu)}\right]+\eta\left[J^{\mu}+(1+\Theta)J_{\nu}u^{\nu}u^{\mu}\right]
\end{equation}
In the limit of pair plasma, the Hall effect disappears as it is proportional to $\mu^{\dagger}$. While the current inertia and thermal electromotive force would be significant. For the ion-electron plasma where $m_+\gg m_-$ such that $\mu\approx 0$ and $\mu^{\dagger}\approx 1$, the equations of motion become:
\begin{equation}
    \partial_{\nu}\left(\omega u^{\mu}u^{\nu}\right)+\partial^{\mu}p=F^{\mu\nu}J_{\nu}
\end{equation}
and
\begin{equation}
    F^{\mu\nu}u_{\nu}-\frac{1}{\frak{q}}\partial^{\mu}p_-=\frac{1}{\frak{q}}F^{\mu\nu}J_{\nu}+\eta\left[J^{\mu}+(1+\Theta)J_{\nu}u^{\nu}u^{\mu}\right]
\end{equation}
In the limit of ion-electron plasma, the Hall effect would be significant while the current inertia and thermal electromotive force disappear.

\section{Electrical resistivity in pair plasma}
\label{sec:resistivity}

Let us consider the simpliest case. In the nearly neutral pair plasma which satisfies $m_+\approx m_-\approx m$ and $n_+\approx n_-\approx n$, positrons and electrons drift in electric field. The equations of motion obey (for non-relativistic approximation):
\begin{equation}
    \begin{aligned}
        \frac{d}{dt}m_+\Vec{v}_+ &= e\Vec{E}+\bar{\nu}_{\rm pe}\left(m_-\Vec{v}_--m_+\Vec{v}_+\right) \\
        \frac{d}{dt}m_-\Vec{v}_- &= -e\Vec{E}-\bar{\nu}_{\rm pe}\left(m_-\Vec{v}_--m_+\Vec{v}_+\right)
    \end{aligned}
    \label{eq:eom_pe}
\end{equation}
where $\bar{\nu}_{\rm pe}$ is the rate of collision between positrons and electrons, quantifying how much the momentum of electrons is transferred to positrons. For the steady state such that $d/dt\sim 0$, combining the equations in Eq.~\eqref{eq:eom_pe}, we get:
\begin{equation}
    \Vec{v}_+-\Vec{v}_- \simeq \frac{e\Vec{E}}{m\bar{\nu}_{\rm pe}}
\end{equation}
The definition of current and the Ohm's law then result in:
\begin{equation}
    \frak{q}\left(\Vec{v}_+-\Vec{v}_-\right) \equiv \Vec{J}=\frac{\Vec{E}}{\eta}~~~\rightarrow~~~
    \eta \simeq \frac{\rho\bar{\nu}_{\rm pe}}{\frak{q}^2}
\end{equation}

\section{Sweet-Parker model in GR}
\label{sec:SP}

In Ref.~\cite{GRMR1}, the calculations of Sweet-Parker model were reorganized to seven steps. Their GR and SR forms are listed below:
\begin{itemize}
\item First, null curl of electric field, namely the spatial components of $\nabla_{\nu}F^{\ast\mu\nu}=0$:
\begin{equation}
    \partial_x E_z=\partial_y E_z=0~~~\xrightleftharpoons[\text{SR form}]{\text{GR form}}~~~
    \tau_{\mu}\partial_{\nu}\sqrt{-g}F^{\ast \mu\nu}=\chi_{\mu}\partial_{\nu}\sqrt{-g}F^{\ast \mu\nu}=0
    \label{eq:step1}
\end{equation}
The $\hat{z}$ component gives a trivial equation ($0=0$).

\item Second, null divergence of magnetic field, namely the time component (along $k^{\mu}$ in GR specifically) of $\nabla_{\nu}F^{\ast\mu\nu}=0$:
\begin{equation}
    \partial_xB_x+\partial_yB_y=0~~~\xrightleftharpoons[\text{SR form}]{\text{GR form}}~~~k_{\mu}\partial_{\nu}\sqrt{-g}F^{\ast \mu\nu}=0
    \label{eq:step2}
\end{equation}

\item Third, Ohm's law upstream and inside the current sheet:
\begin{equation}
    \begin{split}
        {\rm upstream:}&~E_z=-V_yB_x \\
        {\rm current~sheet:}&~E_z=\eta J_z
    \end{split}
    ~~~\xrightleftharpoons[\text{SR form}]{\text{GR form}}~~~
    \begin{split}
        &\zeta_{\mu}F^{\mu\nu}u_{\nu}=0 \\
        &\zeta_{\mu}F^{\mu\nu}u_{\nu}=\eta J^{\mu}\zeta_{\mu}
    \end{split}
    \label{eq:step3}
\end{equation}

\item Fourth, Amp{\`e}re's law surrounding the current sheet, namely the spatial components of $\nabla_{\nu}F^{\mu\nu}=J^{\mu}$:
\begin{equation}
    \partial_xB_y-\partial_yB_x=J_z~~~\xrightleftharpoons[\text{SR form}]{\text{GR form}}~~~
    \zeta_{\mu}\partial_{\nu}\sqrt{-g}F^{\mu\nu}=\sqrt{-g}J^{\mu}\zeta_{\mu}
    \label{eq:step4}
\end{equation}
The other two spatial components provide trivial equations.

\item Fifth, mass conservation:
\begin{equation}
    \partial_x\gamma V_x+\partial_y\gamma V_y=0~~~\xrightleftharpoons[\text{SR form}]{\text{GR form}}~~~\partial_{\mu}\sqrt{-g}u^{\mu}=0
    \label{eq:step5}
\end{equation}

\item Sixth, the $\hat{y}$ component of energy-momentum conservation:
\begin{equation}
    \partial_y\left(\omega\gamma^2V_y^2\right)+\partial_y p=J_zB_x~~~\xrightleftharpoons[\text{SR form}]{\text{GR form}}~~~
    \frac{1}{\sqrt{-g}}\tau_{\mu}\partial_{\nu}\left(\sqrt{-g}\omega u^{\mu}u^{\nu}\right)+
    \tau_{\mu}\partial^{\mu}p=\tau_{\mu}F^{\mu\nu}J_{\nu}-\omega\tau_{\mu}\Gamma^{\mu}_{\kappa\lambda}u^{\kappa}u^{\lambda}
    \label{eq:step6}
\end{equation}

\item Seventh, the $\hat{x}$ component of energy-momentum conservation:
\begin{equation}
    \partial_x\left(\omega\gamma^2v_x^2\right)+\partial_x p=-J_zB_y~~~\xrightleftharpoons[\text{SR form}]{\text{GR form}}~~~
    \frac{1}{\sqrt{-g}}\chi_{\mu}\partial_{\nu}\left(\sqrt{-g}\omega u^{\mu}u^{\nu}\right)+\chi_{\mu}\partial^{\mu}p
    =\chi_{\mu}F^{\mu\nu}J_{\nu}-\omega\chi_{\mu}\Gamma^{\mu}_{\kappa\lambda}u^{\kappa}u^{\lambda}
    \label{eq:step7}
\end{equation}
\end{itemize}

In order to describe the fast magnetic reconnection induced by thermal electromotive force in Sweet-Parker configuration, only the third, sixth and seventh steps the their resultant relations between physical quantities should be reconsidered under the generalized MHD scheme. 

It is inevitable to notice some points and approximations declared in Ref.~\cite{GRMR1}, which may be still useful to handle the equations of generalized MHD in GR:
\begin{itemize}
\item First, derivatives in the equations relate to the variations in local scale. Hence the derivatives acting on the 4-vectors defined in Eq.~\eqref{eq:tetrad} should vanish. For example, the operator $k^{\mu}\partial_{\nu}$ satisfies:
\begin{equation}
    k^{\mu}\partial_{\nu}=\partial_{\nu}k^{\mu}
\end{equation}
when acting on an arbitrary tensor.

\item Second, since the process occurs in the local scale, if we choose, for example, $\chi^{\mu}=\hat{e}_{(\phi)}^{\mu}$, the operator $\chi^{\mu}\partial_{\mu}\sqrt{-g}$ could be approximated as:
\begin{equation}
    \chi^{\mu}\partial_{\mu}\sqrt{-g}=\partial_{\phi}\frac{1}{h_{\phi}}\bigg|_X \alpha h_rh_{\theta}h_{\phi} \simeq \partial_{\phi}\alpha h_rh_{\theta}
\end{equation}
when acting on an arbitrary tensor. This approximation will help to simplify the resultant expressions.

\item Third, as a temporary approximation, we make, for example,
\begin{equation}
    \partial_{r}\alpha h_{\theta}h_{\phi}\simeq 0~~\xrightarrow{\div \alpha h_rh_{\theta}h_{\phi}}~~\frac{1}{h_r}\partial_{r} \simeq 0
    \label{eq:temp}
\end{equation}
It is just neglecting the first order derivatives of $g_{\mu\nu}$. Whether or not the GR forms of equations could map to their SR forms, as shown in Eq.~\eqref{eq:step1}--\eqref{eq:step7}, could be conveniently checked under this temporary approximation. It is worth noticing that the derivation
\begin{equation}
    \partial_{\phi}\alpha h_{\theta}h_{\phi}\simeq 0~~\xrightarrow~~\frac{1}{h_r}\partial_{\phi} \simeq 0
\end{equation}
is always legal because $\alpha$ and $h_i$ do not depend on $\phi$ in Kerr spacetime.

\item Fourth, when substituting the derivatives by finite differences based on Eq.~\eqref{eq:L_delta_GR}, to manage the terms like $f_1\tau^{\mu}\partial_{\mu}f_2$, one could make:
\begin{equation}
    f_1\tau^{\mu}\partial_{\mu}f_2 \simeq f_1\big|_{\varepsilon}\frac{f_2\big|_i-f_2\big|_X}{\delta}
    \label{eq:mean}
\end{equation}
with $\varepsilon$ some point between $i$ and $X$, on which the mean value theorem for finite integral is satisfied.
\end{itemize}

We list below the resultant relations between physical quantities given by Eq.~\eqref{eq:step1}--\eqref{eq:step7} when the magnetic reconnection occurs in the ZAMOs laboratory with current sheet posited azimuthally as Eq.~\eqref{eq:tetrad_z} introduced:
\begin{itemize}
\item Eq.~\eqref{eq:step1} gives:
\begin{equation}
    E_z\simeq const \equiv e_z
    \label{eq:step1-res}
\end{equation}

\item Eq.~\eqref{eq:step2} gives:
\begin{equation}
    \frac{b_1}{b_0}\simeq \mathscr{C}\frac{\delta}{L}\simeq \frac{\delta}{L}
    \label{eq:step2-res}
\end{equation}
with
\begin{equation}
    \mathscr{C}=\frac{\left(\alpha h_{\theta}h_{\phi}\right)\big|_o}{\left(\alpha h_{\theta}h_{\phi}\right)\big|_i}\simeq 1
    \label{eq:C}
\end{equation}

\item Eq.~\eqref{eq:step3} gives:
\begin{equation}
    \begin{split}
        {\rm upstream:}&~\gamma_{\rm in}\left(e_z+v_{\rm in}b_0\right)=0 \\
        {\rm current~sheet:}&~J_z=\frac{E_z}{\eta}\simeq \frac{e_z}{\eta}\equiv j_z \simeq const
    \end{split}
    \label{eq:step3-res}
\end{equation}

\item Eq.~\eqref{eq:step4} gives:
\begin{equation}
    \frac{j_z}{b_0}\simeq -\frac{1}{\delta}\frac{\left(\alpha h_{\theta}h_{\phi}\right)\big|_i}{\left(\alpha h_{\theta}h_{\phi}\right)\big|_{\varepsilon_1}}\simeq -\frac{1}{\delta}
    \label{eq:step4-res}
\end{equation}
where $\varepsilon_1$ is some point between $i$ and $X$. The negative sign tells the direction of current.

\item Eq.~\eqref{eq:step5} gives:
\begin{equation}
    \frac{\gamma_{\rm in}v_{\rm in}}{\gamma_{\rm out}v_{\rm out}}\simeq \mathscr{C}\frac{\delta}{L}\simeq \frac{\delta}{L} 
    \label{eq:step5-res}
\end{equation}

\item Eq.~\eqref{eq:step6} gives:
\begin{equation}
    \frac{p_X}{b_0^2/2}\simeq \frac{\left(\alpha^2h_{\theta}^2h_{\phi}^2\right)\bigg|_i}{\left(\alpha^2h_{\theta}^2h_{\phi}^2\right)\bigg|_{\varepsilon_2}}-
    \frac{\omega}{b_0^2/2}\frac{\ln\alpha\big|_i-\ln\alpha\big|_X}{h_r^2\big|_{\varepsilon_3}}\simeq 1
    \label{eq:step6-res}
\end{equation}
The point $\varepsilon_2$ and $\varepsilon_3$ are two different points between $i$ and $X$.

\item Eq.~\eqref{eq:step7} gives:
\begin{equation}
    v_{\rm out}\simeq \sqrt{\frac{\mathscr{D}\sigma_0}{1+\mathscr{D}\sigma_0}}\simeq v_{\rm A}
    \label{eq:step7-res}
\end{equation}
with:
\begin{equation}
    \mathscr{D}\equiv \frac{1}{2}\left[\frac{\left(\alpha^2h_{\theta}^2h_{\phi}^2\right)\bigg|_i}{\left(\alpha^2h_{\theta}^2h_{\phi}^2\right)\bigg|_{\varepsilon_2}}-
    \frac{\omega}{b_0^2/2}\frac{\ln\alpha\big|_i-\ln\alpha\big|_X}{h_r^2\big|_{\varepsilon_3}} + 
    \frac{\left(\alpha h_{\theta}h_{\phi}\right)\big|_o}{\left(\alpha h_{\theta}h_{\phi}\right)\big|_{\varepsilon_1}}\right] \simeq 1
    \label{eq:D_SP}
\end{equation}
which is different from the factor $\mathscr{D}$ in Eq.~\eqref{eq:D}.

\item Resultantly, the reconnection rate obeys:
\begin{equation}
    R\equiv \frac{v_{\rm in}}{v_{\rm A}}\simeq \mathscr{CD}\frac{\delta}{L}\simeq \mathscr{C}^{-1}\frac{\eta}{\delta v_{\rm A}}\simeq \mathscr{D}^{1/2}S^{-1/2}
    \label{eq:R_SP}
\end{equation}
in the low magnetization limit.
\end{itemize}
Please read Ref.~\cite{GRMR1} for the details.

\bibliographystyle{utphys}
\bibliography{references}

\providecommand{\href}[2]{#2}\begingroup\raggedright\begin{thebibliography}{10}

\bibitem{substorm}
E.~N. {Parker}, ``{The Solar-Flare Phenomenon and the Theory of Reconnection
  and Annihiliation of Magnetic Fields.},''
  \href{http://dx.doi.org/10.1086/190087}{{\em Astrophys. J.} {\bfseries 8}
  (July, 1963) 177}.

\bibitem{2021SSRv..217...66Z}
I.~V. {Zimovets}, J.~A. {McLaughlin}, A.~K. {Srivastava}, D.~Y. {Kolotkov},
  A.~A. {Kuznetsov}, E.~G. {Kupriyanova}, I.~H. {Cho}, A.~R. {Inglis},
  F.~{Reale}, D.~J. {Pascoe}, H.~{Tian}, D.~{Yuan}, D.~{Li}, and Q.~M. {Zhang},
  ``{Quasi-Periodic Pulsations in Solar and Stellar Flares: A Review of
  Underpinning Physical Mechanisms and Their Predicted Observational
  Signatures},'' \href{http://dx.doi.org/10.1007/s11214-021-00840-9}{{\em Space
  Science Reviews} {\bfseries 217} no.~5, (Aug., 2021) 66}.

\bibitem{SP1}
P.~A. {Sweet}, ``{The Neutral Point Theory of Solar Flares},'' in {\em
  Electromagnetic Phenomena in Cosmical Physics}, B.~{Lehnert}, ed., vol.~6,
  p.~123.
\newblock Jan., 1958.

\bibitem{SP2}
E.~N. {Parker}, ``{Sweet's Mechanism for Merging Magnetic Fields in Conducting
  Fluids},'' \href{http://dx.doi.org/10.1029/JZ062i004p00509}{{\em Journal of
  Geophysical Research} {\bfseries 62} no.~4, (Dec., 1957) 509--520}.

\bibitem{Lyubarsky2006}
Y.~E. Lyubarsky, ``{On the relativistic magnetic reconnection},''
  \href{http://dx.doi.org/10.1111/j.1365-2966.2005.08767.x}{{\em Mon. Not. Roy.
  Astron. Soc.} {\bfseries 358} (2005) 113--119},
  \href{http://arxiv.org/abs/astro-ph/0501392}{{\ttfamily
  arXiv:astro-ph/0501392}}.

\bibitem{Comisso:2023ygd}
L.~Comisso and B.~Jiang, ``{Pitch-angle Anisotropy Imprinted by Relativistic
  Magnetic Reconnection},''
  \href{http://dx.doi.org/10.3847/1538-4357/ad1241}{{\em Astrophys. J.}
  {\bfseries 959} no.~2, (2023) 137},
  \href{http://arxiv.org/abs/2310.17560}{{\ttfamily arXiv:2310.17560
  [astro-ph.HE]}}.

\bibitem{Sironi:2014jfa}
L.~Sironi and A.~Spitkovsky, ``{Relativistic Reconnection: an Efficient Source
  of Non-Thermal Particles},''
  \href{http://dx.doi.org/10.1088/2041-8205/783/1/L21}{{\em Astrophys. J.
  Lett.} {\bfseries 783} (2014) L21},
  \href{http://arxiv.org/abs/1401.5471}{{\ttfamily arXiv:1401.5471
  [astro-ph.HE]}}.

\bibitem{yamada2009}
E.~Zweibel and M.~Yamada, ``Magnetic reconnection in astrophysical and
  laboratory plasmas,''
  \href{http://dx.doi.org/10.1146/annurev-astro-082708-101726}{{\em Annual
  Review of Astronomy and Astrophysics - ANNU REV ASTRON ASTROPHYS} {\bfseries
  47} (09, 2009) 291--332}.

\bibitem{1973ApJ...180..247P}
E.~N. {Parker}, ``{The Reconnection Rate of Magnetic Fields},''
  \href{http://dx.doi.org/10.1086/151959}{{\em Astrophysical Journal}
  {\bfseries 180} (Feb., 1973) 247--252}.

\bibitem{Petschek}
H.~E. {Petschek}, ``{Magnetic Field Annihilation},'' in {\em NASA Special
  Publication}, vol.~50, p.~425.
\newblock 1964.

\bibitem{Biskamp1986}
D.~{Biskamp}, ``{Magnetic reconnection via current sheets},''
  \href{http://dx.doi.org/10.1063/1.865670}{{\em Physics of Fluids} {\bfseries
  29} no.~5, (May, 1986) 1520--1531}.

\bibitem{Sato1979}
T.~{Sato} and T.~{Hayashi}, ``{Externally driven magnetic reconnection and a
  powerful magnetic energy converter},''
  \href{http://dx.doi.org/10.1063/1.862721}{{\em Physics of Fluids} {\bfseries
  22} no.~6, (June, 1979) 1189--1202}.

\bibitem{2008PhPl...15d2306D}
J.~F. {Drake}, M.~A. {Shay}, and M.~{Swisdak}, ``{The Hall fields and fast
  magnetic reconnection},'' \href{http://dx.doi.org/10.1063/1.2901194}{{\em
  Physics of Plasmas} {\bfseries 15} no.~4, (Apr., 2008) 042306}.

\bibitem{2006PhPl...13e2119Y}
M.~{Yamada}, Y.~{Ren}, H.~{Ji}, J.~{Breslau}, S.~{Gerhardt}, R.~{Kulsrud}, and
  A.~{Kuritsyn}, ``{Experimental study of two-fluid effects on magnetic
  reconnection in a laboratory plasma with variable collisionality},''
  \href{http://dx.doi.org/10.1063/1.2203950}{{\em Physics of Plasmas}
  {\bfseries 13} no.~5, (May, 2006) 052119}.

\bibitem{Bessho:2005zz}
N.~Bessho and A.~Bhattacharjee, ``{Collisionless Reconnection in an
  Electron-Positron Plasma},''
  \href{http://dx.doi.org/10.1103/PhysRevLett.95.245001}{{\em Phys. Rev. Lett.}
  {\bfseries 95} (2005) 245001}.

\bibitem{Swisdak:2008xn}
M.~Swisdak, Y.~H. Liu, and J.~F. Drake, ``{Development of a Turbulent Outflow
  During Electron-Positron Magnetic Reconnection},''
  \href{http://dx.doi.org/10.1086/588088}{{\em Astrophys. J.} {\bfseries 680}
  (2008) 999}, \href{http://arxiv.org/abs/0803.3415}{{\ttfamily arXiv:0803.3415
  [astro-ph]}}.

\bibitem{2015PhPl...22a0701S}
A.~{Stanier}, A.~N. {Simakov}, L.~{Chac{\'o}n}, and W.~{Daughton}, ``{Fast
  magnetic reconnection with large guide fields},''
  \href{http://dx.doi.org/10.1063/1.4905629}{{\em Physics of Plasmas}
  {\bfseries 22} no.~1, (Jan., 2015) 010701},
  \href{http://arxiv.org/abs/1505.03794}{{\ttfamily arXiv:1505.03794
  [physics.plasm-ph]}}.

\bibitem{Liu2017}
Y.-H. Liu, M.~Hesse, F.~Guo, W.~Daughton, H.~Li, P.~A. Cassak, and M.~A. Shay,
  ``{Why does steady-state magnetic reconnection have a maximum local rate of
  order 0.1?},'' \href{http://dx.doi.org/10.1103/PhysRevLett.118.085101}{{\em
  Phys. Rev. Lett.} {\bfseries 118} no.~8, (2017) 085101},
  \href{http://arxiv.org/abs/1611.07859}{{\ttfamily arXiv:1611.07859
  [physics.plasm-ph]}}.

\bibitem{CA2014}
L.~Comisso and F.~A. Asenjo, ``{Thermal-inertial effects on magnetic
  reconnection in relativistic pair plasmas},''
  \href{http://dx.doi.org/10.1103/PhysRevLett.113.045001}{{\em Phys. Rev.
  Lett.} {\bfseries 113} (2014) 045001},
  \href{http://arxiv.org/abs/1402.1115}{{\ttfamily arXiv:1402.1115
  [physics.plasm-ph]}}.

\bibitem{Koide2009}
S.~Koide, ``{Generalized Relativistic Magnetohydrodynamic Equations for Pair
  and Electron-Ion Plasmas},''
  \href{http://dx.doi.org/10.1088/0004-637X/696/2/2220}{{\em Astrophys. J.}
  {\bfseries 696} (2009) 2220--2233},
  \href{http://arxiv.org/abs/0902.4292}{{\ttfamily arXiv:0902.4292
  [astro-ph.HE]}}. [Erratum: Astrophys.J. 701, 2033 (2009)].

\bibitem{Koide2010}
S.~Koide, ``{Generalized General Relativistic MHD Equations and Distinctive
  Plasma Dynamics around Rotating Black Holes},''
  \href{http://dx.doi.org/10.1088/0004-637X/708/2/1459}{{\em Astrophys. J.}
  {\bfseries 708} (2010) 1459--1474},
  \href{http://arxiv.org/abs/0912.4930}{{\ttfamily arXiv:0912.4930
  [astro-ph.HE]}}.

\bibitem{Jia:2023iup}
H.~Jia, B.~Ripperda, E.~Quataert, C.~J. White, K.~Chatterjee, A.~Philippov, and
  M.~Liska, ``{Millimeter observational signatures of flares in magnetically
  arrested black hole accretion models},''
  \href{http://dx.doi.org/10.1093/mnras/stad2935}{{\em Mon. Not. Roy. Astron.
  Soc.} {\bfseries 526} no.~2, (2023) 2924--2941},
  \href{http://arxiv.org/abs/2301.09014}{{\ttfamily arXiv:2301.09014
  [astro-ph.HE]}}.

\bibitem{Davelaar:2023dhl}
J.~Davelaar, B.~Ripperda, L.~Sironi, A.~A. Philippov, H.~Olivares, O.~Porth,
  B.~v.~d. Berg, T.~Bronzwaer, K.~Chatterjee, and M.~Liska, ``{Synchrotron
  Polarization Signatures of Surface Waves in Supermassive Black Hole Jets},''
  \href{http://dx.doi.org/10.3847/2041-8213/ad0b79}{{\em Astrophys. J. Lett.}
  {\bfseries 959} no.~1, (2023) L3},
  \href{http://arxiv.org/abs/2309.07963}{{\ttfamily arXiv:2309.07963
  [astro-ph.HE]}}.

\bibitem{Aimar2023}
N.~Aimar, A.~Dmytriiev, F.~H. Vincent, I.~E. Mellah, T.~Paumard, G.~Perrin, and
  A.~Zech, ``{Magnetic reconnection plasmoid model for Sagittarius A*
  flares},'' \href{http://dx.doi.org/10.1051/0004-6361/202244936}{{\em Astron.
  Astrophys.} {\bfseries 672} (2023) A62},
  \href{http://arxiv.org/abs/2301.11874}{{\ttfamily arXiv:2301.11874
  [astro-ph.HE]}}.

\bibitem{Ripperda2020}
B.~Ripperda, F.~Bacchini, and A.~Philippov, ``{Magnetic Reconnection and Hot
  Spot Formation in Black Hole Accretion Disks},''
  \href{http://dx.doi.org/10.3847/1538-4357/ababab}{{\em Astrophys. J.}
  {\bfseries 900} no.~2, (2020) 100},
  \href{http://arxiv.org/abs/2003.04330}{{\ttfamily arXiv:2003.04330
  [astro-ph.HE]}}.

\bibitem{Yuan2024-1}
H.~Yang, F.~Yuan, H.~Li, Y.~Mizuno, F.~Guo, R.~Lu, L.~C. Ho, X.~Lin, A.~A.
  Zdziarski, and J.~Wang, ``{Modeling the inner part of the jet in M87:
  Confronting jet morphology with theory},''
  \href{http://dx.doi.org/10.1126/sciadv.adn3544}{{\em Sci. Adv.} {\bfseries
  10} no.~12, (2024) adn3544},
  \href{http://arxiv.org/abs/2403.15950}{{\ttfamily arXiv:2403.15950
  [astro-ph.HE]}}.

\bibitem{Yuan2024-2}
L.~Xi and Y.~Feng, ``{Revisiting flares in Sagittarius A* based on general
  relativistic magnetohydrodynamic numerical simulations of black hole
  accretion},'' \href{http://dx.doi.org/10.1093/mnras/stae1357}{{\em Mon. Not.
  Roy. Astron. Soc.} {\bfseries 531} no.~3, (2024) 3136--3150},
  \href{http://arxiv.org/abs/2405.17408}{{\ttfamily arXiv:2405.17408
  [astro-ph.HE]}}.

\bibitem{KA2008}
S.~Koide and K.~Arai, ``{Energy Extraction from a Rotating Black Hole by
  Magnetic Reconnection in Ergosphere},''
  \href{http://dx.doi.org/10.1086/589497}{{\em Astrophys. J.} {\bfseries 682}
  (2008) 1124}, \href{http://arxiv.org/abs/0805.0044}{{\ttfamily
  arXiv:0805.0044 [astro-ph]}}.

\bibitem{CA2021}
L.~Comisso and F.~A. Asenjo, ``{Magnetic Reconnection as a Mechanism for Energy
  Extraction from Rotating Black Holes},''
  \href{http://dx.doi.org/10.1103/PhysRevD.103.023014}{{\em Phys. Rev. D}
  {\bfseries 103} no.~2, (2021) 023014},
  \href{http://arxiv.org/abs/2012.00879}{{\ttfamily arXiv:2012.00879
  [astro-ph.HE]}}.

\bibitem{Work0}
B.~Chen, Y.~Hou, J.~Li, and Y.~Shen, ``{Energy extraction from a Kerr black
  hole via magnetic reconnection within the plunging region},''
  \href{http://dx.doi.org/10.1103/PhysRevD.110.063003}{{\em Phys. Rev. D}
  {\bfseries 110} no.~6, (2024) 063003},
  \href{http://arxiv.org/abs/2405.11488}{{\ttfamily arXiv:2405.11488 [gr-qc]}}.

\bibitem{Work1}
Y.~Shen, H.-Y. YuChih, and B.~Chen, ``{Energy extraction from a rotating black
  hole via magnetic reconnection: the plunging bulk plasma and orientation
  angle},'' \href{http://arxiv.org/abs/2409.07345}{{\ttfamily arXiv:2409.07345
  [gr-qc]}}.

\bibitem{CA2017}
F.~A. Asenjo and L.~Comisso, ``{Relativistic Magnetic Reconnection in Kerr
  Spacetime},'' \href{http://dx.doi.org/10.1103/PhysRevLett.118.055101}{{\em
  Phys. Rev. Lett.} {\bfseries 118} no.~5, (2017) 055101},
  \href{http://arxiv.org/abs/1701.03669}{{\ttfamily arXiv:1701.03669
  [astro-ph.HE]}}.

\bibitem{Fan:2024fcy}
Z.-Y. Fan, Y.~Li, F.~Zhou, and M.~Guo, ``{Fast magnetic reconnection in Kerr
  spacetime},'' \href{http://arxiv.org/abs/2409.05434}{{\ttfamily
  arXiv:2409.05434 [astro-ph.HE]}}.

\bibitem{CA2018}
L.~Comisso and F.~A. Asenjo, ``{Collisionless Magnetic Reconnection in Curved
  Spacetime and the Effect of Black Hole Rotation},''
  \href{http://dx.doi.org/10.1103/PhysRevD.97.043007}{{\em Phys. Rev. D}
  {\bfseries 97} no.~4, (2018) 043007},
  \href{http://arxiv.org/abs/1801.06174}{{\ttfamily arXiv:1801.06174
  [astro-ph.HE]}}.

\bibitem{Fan:2024rsa}
Z.-Y. Fan, F.~Zhou, Y.~Li, M.~Guo, and B.~Chen, ``{Magnetic reconnection under
  centrifugal and gravitational electromotive forces},''
  \href{http://arxiv.org/abs/2411.19491}{{\ttfamily arXiv:2411.19491
  [astro-ph.HE]}}.

\bibitem{GRMR1}
Y.~Shen, ``{How to describe the Sweet-Parker model in general relativity},''
  \href{http://arxiv.org/abs/2409.16596}{{\ttfamily arXiv:2409.16596 [gr-qc]}}.

\bibitem{BirnBook}
J.~{Birn} and E.~R. {Priest},
  \href{http://dx.doi.org/10.1017/CBO9780511536151}{{\em Reconnection of
  Magnetic Fields: Magnetohydrodynamics and Collisionless Theory and
  Observations}}.
\newblock Cambridge University Press, 2007.

\bibitem{Vincent:2022fwj}
F.~H. Vincent, S.~E. Gralla, A.~Lupsasca, and M.~Wielgus, ``{Images and photon
  ring signatures of thick disks around black holes},''
  \href{http://dx.doi.org/10.1051/0004-6361/202244339}{{\em Astron. Astrophys.}
  {\bfseries 667} (2022) A170},
  \href{http://arxiv.org/abs/2206.12066}{{\ttfamily arXiv:2206.12066
  [astro-ph.HE]}}.

\bibitem{Hou:2023bep}
Y.~Hou, Z.~Zhang, M.~Guo, and B.~Chen, ``{A new analytical model of
  magnetofluids surrounding rotating black holes},''
  \href{http://dx.doi.org/10.1088/1475-7516/2024/02/030}{{\em JCAP} {\bfseries
  02} (2024) 030}, \href{http://arxiv.org/abs/2309.13304}{{\ttfamily
  arXiv:2309.13304 [gr-qc]}}.

\bibitem{Zhang:2024lsf}
Z.~Zhang, Y.~Hou, M.~Guo, and B.~Chen, ``{Imaging thick accretion disks and
  jets surrounding black holes},''
  \href{http://dx.doi.org/10.1088/1475-7516/2024/05/032}{{\em JCAP} {\bfseries
  05} (2024) 032}, \href{http://arxiv.org/abs/2401.14794}{{\ttfamily
  arXiv:2401.14794 [astro-ph.HE]}}.

\bibitem{MacDonald:1982zz}
D.~MacDonald and K.~S. Thorne, ``{Black-hole electrodynamics - an
  absolute-space/universal-time formulation},'' {\em Mon. Not. Roy. Astron.
  Soc.} {\bfseries 198} (1982) 345--383.

\end{thebibliography}\endgroup

\end{document}